\documentstyle[11pt,aaspp4]{article}
 
\slugcomment{To appear in ApJ, 493, Feb 1, 1998}
\lefthead{Hunter et al.}
\righthead{Star Formation in Irregular Galaxies}
 
\def\x{\enspace}
\def\xx{\enspace\enspace}
\def\xxx{\enspace\enspace\enspace}
\def\xxxx{\enspace\enspace\enspace\enspace}

\def\et{et al.}
\def\kms{km s$^{-1}$}
\def\ha{H$\alpha$}

\def\solar{\ifmmode_{\mathord\odot}\;\else$_{\mathord\odot}\;$\fi}
\def\sigcrit{$\Sigma_c$}
\def\siggas{$\Sigma_g$}
\def\sigrat{$\Sigma_g/\Sigma_c$}
\def\sigeff{$\Sigma_{c,2f}$}
\def\sigrateff{$\Sigma_g/\Sigma_{c,2f}$}
\def\rhii{$R_{H\alpha}$}
\def\r25{$R_{25}$}

\begin{document}
 
\title{The Relationship Between Gas, Stars, and Star Formation
in Irregular Galaxies: A Test of Simple Models}
 
\author{Deidre A.\ Hunter}
\affil{Lowell Observatory, 1400 West Mars Hill Road, Flagstaff, Arizona 86001
USA; \\dah@lowell.edu}
 
\author{Bruce G.\ Elmegreen}
\affil{IBM T.\ J.\ Watson Research Center, PO Box 218, Yorktown Heights,
New York 10598 USA; \\bge@watson.ibm.com}
 
\and
 
\author{Aomawa L.\ Baker}
\affil{Department of Earth, Atmospheric, and Planetary Sciences,
Massachusetts Institute of Technology, Cambridge, Massachusetts,
02139 USA; \\guinan@mit.edu}
 
\begin{abstract}
 
Irregular galaxies are a unique test of models for the physical laws regulating
star formation because of their lack of spiral density waves and
rotational shear.  Here we explore various instability models for the
onset of star formation in irregular galaxies.	If the gas is
unstable, clouds and eventually stars can form, and so these models
should predict where star formation occurs.  Critical gas densities
were calculated for gravitational instabilities in two models, one
with a thin, pure-gas disk ($\Sigma_c$), and another with a thick disk
composed of gas and a star-like fluid ($\Sigma_{c,2f}$).  We also
calculated the stability properties of three dimensional systems
including dark matter, considered the thermal state of the gas,
and used a modified threshold column density written in terms of the local
rate of shear instead of the epicyclic frequency.
The model predictions were compared to the azimuthally-averaged
present day star formation activity traced by the \ha\ surface
brightness, and to the 1 Gyr-integrated star formation activity
represented by the stellar surface brightness.
 
We find that the ratio of the observed gas density to the critical gas
density, \sigrat, is lower by a factor of $\sim2$ in most of the Im
galaxies than it is in spiral galaxies, at both the intermediate
radii where \sigrat\ is highest, and 
at the outer radii where star formation ends.  
We also find
that, although star formation in irregulars usually occurs 
at intermediate radii
where \sigrat\ is highest, this activity often ends
before \sigrat\ drops significantly in the outer regions, and
it remains high in the inner regions
where \sigrat\ is often low. 
There are
also no correlations between the peak, average, or edge values
of \sigrat\ and the integrated star
formation rates in irregulars.  
These results suggest that
\sigrat\ does not trace star formation with the
same detail in irregular galaxies as it appears to trace it in giant
spiral galaxies.
The low value of $\alpha$ also implies that
either the gas in irregulars is more stable
than it is in spirals, or $\Sigma_c$ 
is not a good threshold for star-forming instabilities. 

Dark matter in the disks of irregulars makes the gas
more unstable, but stars do the same for
the disks of
spirals, leaving the ratio of the two $\alpha$-values about the same.
Moreover, 
the instability parameter with dark matter
still does not follow the star formation activity in irregulars. 
The thermal model suggests that irregulars have
difficulty in sustaining a cool, dense gas phase, and it too fails to
predict where star formation occurs.
An alternative model in which cloud formation involves a competition
between self-gravity and shear, rather than an instability in the
usual sense,
is more successful in defining the threshold for star formation,
but it does not predict where star formation ends either. 
The failure of these models
suggests that processes other than spontaneous instabilities are 
important for star formation in irregular galaxies.  

The role of
\sigrat\ in {\it spiral} galaxies is also questioned. The observed
sensitivity of the star formation rate to \sigrat\ may be strongly
dependent on instabilities specific to spiral arms and not general
instabilities of the type for which \sigrat\ was originally derived.
In that case, large-scale star formation may end in the outer
disks of spirals because the stellar 
density waves end there, at the outer Lindblad resonance. 
 
The only azimuthally-averaged quantity that correlates with the
current star formation activity in irregulars is the stellar surface
density.  A causal connection is possible if stellar energy input to
the interstellar medium acts as a feedback process to star formation.
If this process played a key role in initiating star formation in
irregulars from the beginning, then it could explain why irregular
galaxies began their evolution slowly compared to larger disk systems
with spiral arms.
 
\end{abstract}
 

\indent received 25 June 1997; accepted 28 August 1997;
To appear in ApJ, 493, Feb 1, 1998
 
\section{Introduction}
 
Why do disk galaxies often have gas extending far beyond the optical
Holmberg radius?  Stars form from gas, but few stars form in the outer
gas disks surrounding galaxies.  Intuitively this makes sense:  as the
density of gas drops with increasing distance from a galaxy's center,
there should come a point where clouds do not form to make stars.  One
way to account for this is with a threshold density that is either
constant (Madore, van den Bergh, \& Rogstad 1974; Mouschovias 1981;
Seiden 1983; Gallagher \& Hunter 1984; Hunter \& Gallagher 1986;
Guiderdoni 1987; Skillman 1987; Buat, Deharveng, \& Donas 1989) or
dependent on local conditions (Quirk 1972; Kennicutt 1989; Elmegreen \&
Parravano 1994).  A threshold density that depends on local conditions
is in keeping with observations that show historically constant star
formation activity as a function of radius in irregulars (Hunter \&
Gallagher 1985; Hunter, Gallagher, \& Rautenkranz 1982) and spirals
(Hodge \& Kennicutt 1983).  However, the physical origin of such a
threshold is not well understood.
 
In a thin, differentially-rotating disk composed of pure gas (Safronov
1960) or pure stars (Toomre 1964), there is a critical column density
\sigcrit\ above which the disk is unstable to ring-like perturbations in
the radial direction.  Such ring instabilities are not actually observed
in galaxies, even for column densities $\Sigma>\Sigma_c$ (galaxies form
rings by other processes; see Byrd et al. 1994), but there is an
empirical dependence for star formation that involves this threshold
anyway.  Kennicutt (1989) calculated the critical gas density in the
thin rotating disk model (Quirk 1972) for a sample of normal Sc spiral
galaxies.  He found that the ratio of observed gas density \siggas\ to
critical gas density \sigcrit\ typically exceeds 1 at mid-radius in the
optical disk, and then falls with increasing distance from the center of
the galaxy; star formation was not detected beyond the point where
\sigrat\ dropped below 0.7$\pm$0.2. He concluded that in Sc galaxies the
place where \sigrat$<$0.7 is too stable to form stars; interior to this,
the gas density exceeds the local threshold for star formation.  The
first indication of this effect was in Zasov \& Simakov
(1988); the general concept that galaxy edges result from
below-threshold \sigrat\ dates back to Fall \& Efstathiou (1980).
 
The utility of \sigrat\ as a criterion for star formation is remarkable
considering all of the important physical processes that affect the value
of \sigcrit.  One complication is that realistically thick disks are
predicted to have higher column density thresholds by a factor of $\sim
1.5$ compared to thin disks because the component of the self-
gravitational force parallel to the plane is diluted for the off-plane
mass (Romeo 1992).  Another problem is that the threshold for {\it
radial} instabilities in the gas should be higher still when an
azimuthal magnetic field is included, by a factor of
$(v_A^2+c^2)^{1/2}/c\sim2^{1/2}$ 
for Alfven speed $v_A$ and turbulent gas speed $c$
(Elmegreen 1992).
 
Other considerations lower the threshold.  Energy dissipation that gives
an effective ratio of specific heats $\gamma<1$ for equation of state
$P\propto\rho^{\gamma}$ ($P$= pressure, $\rho$= density) leads to an
effective turbulent speed $\gamma^{1/2}c$.  Because $\Sigma_c\propto
\kappa c/(\pi G)$ for epicyclic frequency $\kappa$, this decreases
$\Sigma_c$ by $\gamma^{1/2}$ when there is no magnetic field and
increases it by only $(v_A^2+\gamma c^2)^{1/2}/c$ when there is a
magnetic field (Elmegreen 1992).  Viscosity removes the column density
threshold for radial collapse entirely:  viscous gas is always unstable,
but the growth approaches the long viscous time when the column density
is below the usual threshold (Elmegreen 1995a; Gammie 1996).  The
presence of stars also lowers the threshold in a combined star+gas
instability (Jog \& Solomon 1984).
 
A more important consideration is that cloud formation is 
three-dimensional, 
so star formation should require both radial and azimuthal
instabilities, or the formation of spiral arms with parallel
instabilities in them (``beads on a string'').  Stability against the
initial growth of spiral or azimuthal perturbations (the ``swing
amplifier'') requires a value of the column density that is lower by a
factor of 2 or more compared to $\Sigma_c$ (Toomre 1981; Athanassoula
1984).  There is no sharp threshold for azimuthal growth either, and it
is always transient for a polytropic, non-magnetic disk because
perturbations eventually shear away.  An azimuthal magnetic field can
change this latter situation because the growing condensations lose
angular momentum as the field lines twist up in response to the Coriolis
force, and this angular momentum loss causes them to continue to grow,
especially if the gas radiates and loses energy (Elmegreen 1987, 1991).
Still, there is no sharp density threshold for this process, unlike
radial instabilities in a thin disk, because purely azimuthal
instabilities generally have no sharp threshold, and because the
gravitational instability is always accompanied by the Parker (1966)
instability, in which the field lines buckle perpendicular to the plane
at a rate that is independent of density (Elmegreen 1982, 1991).
 
Two recent studies point to a possible origin of the
observed density threshold for star formation when there are spiral
arms.  One considers the azimuthal or swing-amplifier instability
discussed above, including magnetic forces, the Parker instability, and
a time-dependent equation for heating and cooling (Elmegreen 1991).  The
result shows that, whereas the {\it primary} instability leading to
transient spiral arms has no sharp density threshold, a {\it secondary}
instability inside these arms, leading to discrete, three-dimensional
clouds, does have a sharp threshold.  The reason for this is that the
secondary instability contains a strong component in the radial
direction, and so is a remnant of the pure-ring instability discussed
above.  It follows that regions with subcritical column densities can
have transient gas spiral arms but no significant collapse to stars.  This
may explain how outer galaxy disks can have normal-looking HI spiral arms but
no significant star formation (e.g., Boulanger \& Viallefond 1992).
 
A second study considers the analogous collapse of magnetic gas that
has been compressed by a steady spiral density wave (Elmegreen 1994a).
The collapse occurs along the arm, in the direction parallel to
the spiral arm.  This leads to the commonly observed ``beads on a string''
phenomenon of star formation in grand-design spiral arms.  Again there
is a column density threshold similar to that for the pure-ring
collapse, but now the result follows more from timing than from an
absolute stability condition.  This is because the parallel
perturbations always begin to grow regardless of column density (a long
thin filament like a dustlane is always gravitationally unstable to
perturbations along its length), but the growth is significant, leading
to giant self-gravitating clouds and star formation, only if it is fast
enough to occur {\it while} the material is compressed.  That is, if the
flow-through time in the arm is short compared to the collapse time, the
gas will not collapse.  The condition for sufficiently rapid unstable
growth is that the average column density in the arms and interarms has
to exceed a threshold that is about the same as that for the radial
instability without a magnetic field, which is just the empirical value
observed for star formation.
 
Other star formation processes not involving spiral arms have been
proposed to give a \sigcrit\ threshold too.  For example, sequentially
triggered (star-induced) star formation works primarily where
\sigrat$>$1, because then the swept-up shell or ring has time to
collapse and form stars before it shears away (Elmegreen 1994b).  The
same is true for turbulence-initiated star formation, which works best
when the turbulent medium is close to the threshold for gravitational
collapse (Elmegreen 1993a).
In addition, cloud formation by the gradual condensation of gas
along magnetic field lines in the presence of shear
gives a threshold similar to \sigcrit\ in regions with flat rotation curves,
but different by a factor of $\sim2$ in regions with low shear (Elmegreen 1993b).
Irregular galaxies may be dominated by such
spiral-free mechanisms (Hunter \& Plummer 1996).  Some contain giant HI holes or
shells with peripheral star formation (for the LMC, see Dopita,
Mathewson, \& Ford 1985; for HoII, see Puche et al. 1992), there is
usually more turbulence relative to rotation in these galaxies (Tully
\et\ 1978; Carignan \et\ 1990; Young \& Lo 1996), and
the rate of shear is generally low.
 
These considerations support what seems to have become a standard model
dating back to Goldreich \& Lynden-Bell (1965), that a wide
variety of processes all dependent on \sigrat\ conspire to initiate most
star formation in galaxies.  These processes may not actually determine
the star formation rate, because that also depends on cloud destruction
and turbulence stirring, but they ensure that there is some kind of feedback
in which the star formation rate controls itself, possibly by regulating
the velocity dispersion that contributes to \sigcrit.  Then in a region
where any particular cloud formation mechanism is absent, as
in an irregular or flocculent galaxy that has no prominent spiral waves,
other mechanisms become relatively more important and take over,
preserving the same overall star formation rate.
This model has been applied to giant spiral galaxies
(Kennicutt 1989;
Caldwell et al. 1992), elliptical galaxies (Vader \& Vigroux 1991),
low surface brightness galaxies (van der Hulst
et al. 1993), and regions of intense star formation (Shlosman
\& Begelman 1989; Elmegreen 1994c).
 
Here we investigate \sigrat\ and other
thresholds for star formation in a sample of irregular galaxies, looking
for trends that can serve as clues to the star formation processes
in these systems.
We have examined galaxies for which HI data exist in the literature with
adequate spatial resolution such that the optical galaxy is
significantly larger than the HI beamwidth.  The observed gas density
relative to the critical gas density is compared to the current ($<$10
Myr) star formation activity determined from \ha\ images and to the
stellar surface density, representing star formation integrated over
$\sim$1 Gyr (Gallagher, Hunter, \& Tutukov 1984).  Broad-band B or V
surface photometry is used to determine the stellar surface density.
 
This investigation began in a previous paper, where Hunter \& Plummer
(1996) examined the radially averaged gas densities in one irregular
galaxy, Sextans A. They found that the ratio of the observed gas density
to the critical gas density \sigrat\ predicted by the thin rotating gas
disk model is everywhere lower than it is in the Sc galaxies examined by
Kennicutt (1989), by roughly a factor of two\footnote{Figure 4 of Hunter
\& Plummer (1996) shows \sigcrit\ to a radius R of 1.4 kpc for the east
side of the galaxy and 1.7 kpc for the west side.  However, $\kappa$
becomes imaginary beyond R$=$1.2 kpc because the rotation curve turns
over and drops steeply there.  Thus, \sigcrit\ should not have been
plotted for R$>$1.2 kpc.
Here we interpolate over the region of the steep drop in the
rotation curve of Sextans A to make $\kappa$ real.}.
This seems to suggest that the gas in
Sextans A is more stable than it is in spiral galaxies, making it harder
for Sextans A to form clouds and stars.  It should be more stable still
because Sextans A probably has a thick disk, as do other irregular
galaxies (Hodge \& Hitchcock 1966; van den Bergh 1988; Puche \et\ 1992),
and this increases $\Sigma_c$, as discussed above.  Yet stars are
forming at a normal rate in Sextan A, as if \sigrat\ is not applicable.
Other subcritical regions with normal star formation include the inner
disks of the spiral galaxies M33 (Kennicutt 1989;
Wilson, Scoville, \& Rice 1991) and NGC 2403 (Thornley \& Wilson 1995).
 
Even though the absolute value of \sigrat\ is low in Sextans A, the
radial trends in this quantity still follow the star formation activity.
For example, the present-day star formation traced by \ha\ emission is
generally higher at radii where \sigrat\ is higher (Hunter \& Plummer
1996).  Van Zee \et\ (1996, 1997) also found this trend for a sample
of irregular galaxies.
Taylor \et\ (1994) showed, in addition, that HII regions
in HII galaxies fall within the critical HI column density contour
predicted by the theory.  Also, van der Hulst \et\ (1993) found that
\sigrat\ is highest where H$\alpha$ is strongest in low surface
brightness spirals, even though in the centers
of these galaxies values of \sigrat\ are lower
than for bright Sc galaxies.
We consider similar trends in the irregular galaxy sample
studied here.

\section{Galaxy Samples}
 
The galaxies are listed in Table \ref{tabgal} along with a few basic
parameters.  The references for the HI data are given in Table
\ref{tabref}.  We have two galaxy samples. The first
consists of relatively nearby, normal Im-type
galaxies.
DDO 154 is notable for its extended HI relative to the
optical galaxy; the HI reaches to five times the Holmberg radius at a
column density of $1\times10^{19}$ cm$^{-2}$ (Carignan \& Beaulieu
1989).  DDO 155 ($=$GR8) is a tiny galaxy with an integrated B
luminosity that is not much greater than that of a large star cluster.
Carignan, Beaulieu, \& Freeman (1990) found that half of the gas in DDO
155 is located in clouds just outside the optical image and that the
rotation axis is parallel to the optical major axis.  DDO 50
($=$Holmberg II) is an irregular galaxy in the M81 group;
Puche \et\ (1992) found that the HI distribution is full of holes, some
as large as 1.5 kpc in diameter, that have been carved in the disk
presumably by concentrations of massive stars.  IC1613 is a Local Group
irregular containing numerous OB associations (Hodge 1978).  To these
galaxies we add Sextans A from a previous study (Hunter \& Plummer
1996).
 
The second galaxy sample consists of more distant galaxies classified
as Im and Sm in the
surveys for very low surface brightness objects conducted by Schombert
\et\ (1988, 1992).
This sample is
an extension of the nearby Im class
that was considered previously to larger 
galaxies that also have no prominant spiral arms. 
The resolution of the HI relative to the optical size
is generally lower than for the nearby Im galaxies.
Most of the distant galaxies 
have central surface brightnesses that are comparable to
or a little fainter than those of
the nearby Im sample, while the disk scale lengths
are always larger 
for the distant galaxies (de Blok \et\ 1995; McGaugh, Schombert,
\& Bothun 1995; Carignan \& Beaulieu 1989; Carignan, Beauleiu, \& Freeman
1990; Broeils 1992). The luminosities of the distant galaxies
range from 1.2 magnitude fainter
to 1.3 magnitudes brighter than that of the LMC.
Mass models for four of the distant galaxies 
and two of the nearby Im galaxies indicate that
many of the distant galaxies
have masses that are greater than those of typical irregular
galaxies (de Blok \& McGaugh 1997). For example, the stellar and dark
matter masses and central mass density of F565-V2,
and the stellar mass and central density
of F583-1, are comparable to those of irregular galaxies, but
in the cases of F563-1 and F563-V2, they 
are significantly higher.
Thus, at least one of the distant galaxies has properties that
are similar to those of irregulars, but others are more massive
and larger.
Nevertheless, all of the distant galaxies 
have weak or no spiral arms,
and they
provide a good comparison to the normal spirals examined
by Kennicutt (1989).
 
There is some confusion in the literature about what constitutes a
``dwarf'' galaxy (see, for example, the review by Binggeli 1994).
The term was used long before any quantitative definition
was applied (Reeves 1997). 
Dwarfs have been defined in terms of 
surface brightness (Hunter \& Gallagher
1986), intrinsic diameter (Pildis, Schombert, \& Eder 1997), and
absolute brightness, with 
M$_B$ or M$_V$ thresholds ranging 
from $-$15 to $-$17 mag
(Hodge 1971, Tammann 1980, Schombert \et 1995).
Clearly, 
not all irregular galaxies are dwarfs (Reeves 1952),
and not all dwarfs are irregulars
(e.g., Schombert \et\ 1995). For example, according to the M$_B$ cuffoffs,
the Large Magellanic Cloud, which typifies 
the Im class, is not a dwarf galaxy.
In this paper we are examining ``irregular'' galaxies, not specifically
``dwarf'' galaxies. It happens that most, but not all, of
the galaxies in our nearby
Im sample would be considered dwarfs, while most, but not all, of the
galaxies in the distant sample would not (although two of those
that would not make the M$_B$ cutoff appear in the Pildis \et\ [1997]
catalog of gas-rich dwarfs based on a size criterion---F563-V1 and
F563-V2). The point is that we are examining the star formation
process in {\it irregular} galaxies, which generally
do not have strong spiral density waves.
 
\section{The Data}
 
\subsection{\ha\ Images}
 
\ha\ images of DDO 168 and IC1613 have already been presented by Hunter,
Hawley, \& Gallagher (1993).  Others were imaged with a TI CCD on the
Perkins 1.8 m telescope at Lowell Observatory using redshifted \ha\
filters with FWHMs of 30 \AA.  A 95 \AA-wide filter centered at 6440
\AA\ was used to determine the stellar continuum which was subtracted
from the \ha\ image to leave only nebular emission.  The nebular flux
calibration used HII regions whose \ha\ fluxes through a particular
aperture size were known from spectrophotometry and spectrophotometric
standard stars.
 
Broad-band V or the red continuum images were used to determine by eye
the isophotal center of the galaxy.  For IC1613 we did not image the
entire galaxy.  However, comparisons with the HII region surveys of
Hodge, Lee, \& Gurwell (1990) and Price, Mason, \& Gullixson (1990)
suggest that we did not miss very much \ha\ flux.  We have used the
lists of HII regions in these references to add to the photometry those
HII regions that were not included in our image.  To determine the
center of the galaxy on our \ha\ image, we compared the \ha\ to a
continuum image shown by Sandage (1971), a contour plot shown by Hodge
\et\ (1990), and the kinematic center determined by Lake \& Skillman
(1989).
 
Elliptical photometry was performed on the \ha\ images using position
angles and ellipticities given by the HI references.  The values are
given in Table~\ref{tabref}.  From this we calculated the average \ha\
surface brightness in each elliptical annulus.  The L$_{H\alpha}$ are
corrected for extinction using the Milky Way foreground value given in
Table \ref{tabgal} plus internal extinction.  When a measure of the
internal extinction was not available, we assumed an E(B$-$V) of 0.05.
We used the Schild (1977) reddening function.
 
The total galactic star formation rate was determined from the
integrated L$_{H\alpha}$ and the formula given by Hunter \& Gallagher
(1986).  This assumes a Salpeter (1955) initial stellar mass function
from 0.1 to 100 M\solar.  Contributions to the star formation rate from
stars below $\sim0.5$ M$_\odot$, where the IMF seems to level off
(Kroupa 1995), are overestimated with the Salpeter initial mass function,
but the absolute values of the star formation rates are not important;
we discuss only radial trends and comparisons to other galaxies where
the same IMF was used.  To remove galaxy size variations, the star
formation rate is normalized to the area of the galaxy within the
surface brightness contour 25 B-magnitudes arcsec$^{-2}$.  The results
are given in Table \ref{tabgas}.  We also list the radius \rhii, which
is the largest radius at which \ha\ emission is detected.

\subsection{Broad-band Images}
 
We used broad-band surface photometry that was available in the
literature for several galaxies.  The references are in Table
\ref{tabref}.  For DDO 50 and DDO 105, we obtained V-band images with
the Perkins 1.8 m telescope and a TI CCD on loan from the US Naval
Observatory.  For color information, U$-$B or B$-$V, we used
azimuthally-averaged data for Sextans A, DDO 154, and DDO 155 using the
references given in the table.  For the rest we used integrated color
measurements from de Vaucouleurs \et\ (1991, $=$RC3) which were assumed
to apply to the galaxy at all radii.  Because color gradients in
irregular galaxies are small, this assumption is a reasonable
approximation.
 
\subsection{Gas Densities}
 
Observed HI densities, deprojected to the plane of the galaxy, were
taken from the references given in Table \ref{tabref}.  We corrected for
the presence of He which we took to be 1.34 times the HI content.
Including molecular gas is far more difficult; few irregular galaxies
have been adequately mapped, so the spatial distribution of H$_2$ is not
generally known.  Instead, we took a globally averaged
M$_{H_2}$/M$_{HI}$ ratio and simply multiplied the HI everywhere by this
in order to statistically include H$_2$.  For the M$_{H_2}$/M$_{HI}$
ratio we used an average from the SMC, where molecular gas is 0.07\% of
the HI gas content (Rubio \et\ 1991), and the LMC, where the ratio is
30\% (Cohen \et\ 1988).  This correction for the presence of H$_2$ is
clearly unsatisfactory, but at present there is no better information on
these galaxies.  Thus, we took \siggas\ to be 1.52$\times$$\Sigma_{HI}$.
 
\section{Comparison with the Rotating Disk Model}
 
\subsection{Single-fluid Thin Disk Model}
 
First we consider the model of a single-fluid, thin disk.  To calculate
the critical gas density of this model we used the formalism of
Kennicutt (1989) in which $\Sigma_c=\alpha\kappa c/\left(3.36G\right)$,
where $\kappa$ is the epicyclic frequency and $c$ is the velocity
dispersion of the gas.  The free parameter $\alpha$ is assumed to have a
value of one initially, but Kennicutt determined it from the ratio
\sigrat\ at the radius of the last detectable HII region.  He measured a
value of $\alpha$$=$0.7$\pm$0.2 for Sc spiral galaxies.  We will also
assume $\alpha$ to be one
initially
here and determine \sigrat\ at the radius
where H$\alpha$ emission is no longer detected for our sample of
galaxies for comparison to Kennicutt's result.  The values of $c$,
listed in Table~\ref{tabref}, were taken from the HI references for
IC1613, DDO 155, DDO 50, and Sextans A; it was assumed to be 9 \kms\ for
the other galaxies.  For DDO 155, \sigcrit\ could only be computed for
the inner part of the galaxy because the outer part (R$\geq$500 pc) is
supported by random motions rather than rotational (Carignan \et\ 1990).
 
Values of $\alpha$ and maximum \sigrat\ are given in Table \ref{tabgas}
for each galaxy.  We have also computed an area-weighted average
\sigrat\ for each galaxy within the radius \rhii.  These are listed in
Table \ref{tabgas}.
 
\subsubsection{Nearby Im Galaxy Sample}
 
\begin{figure}
\vspace{7.0in}
\includegraphics{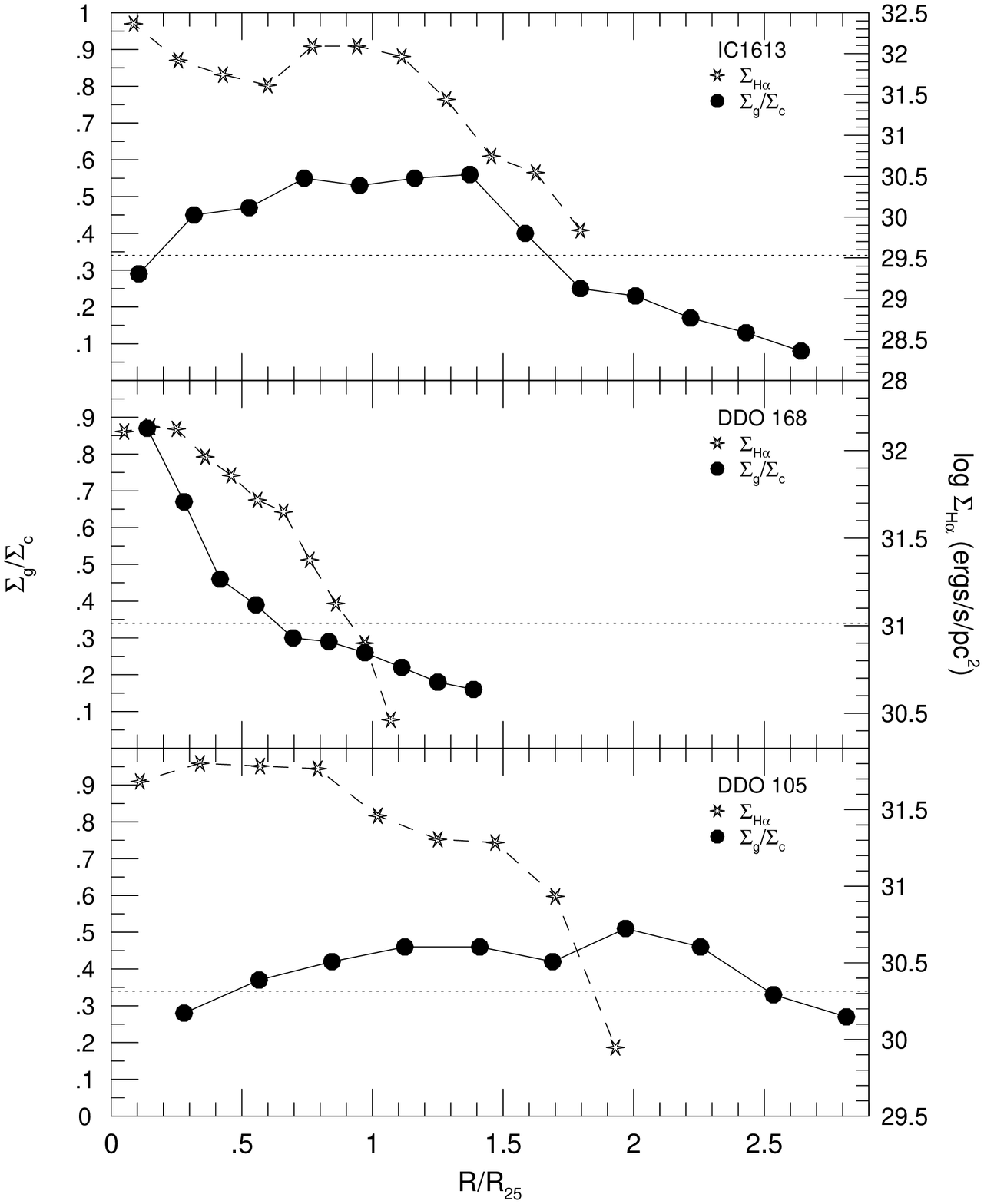}
\includegraphics{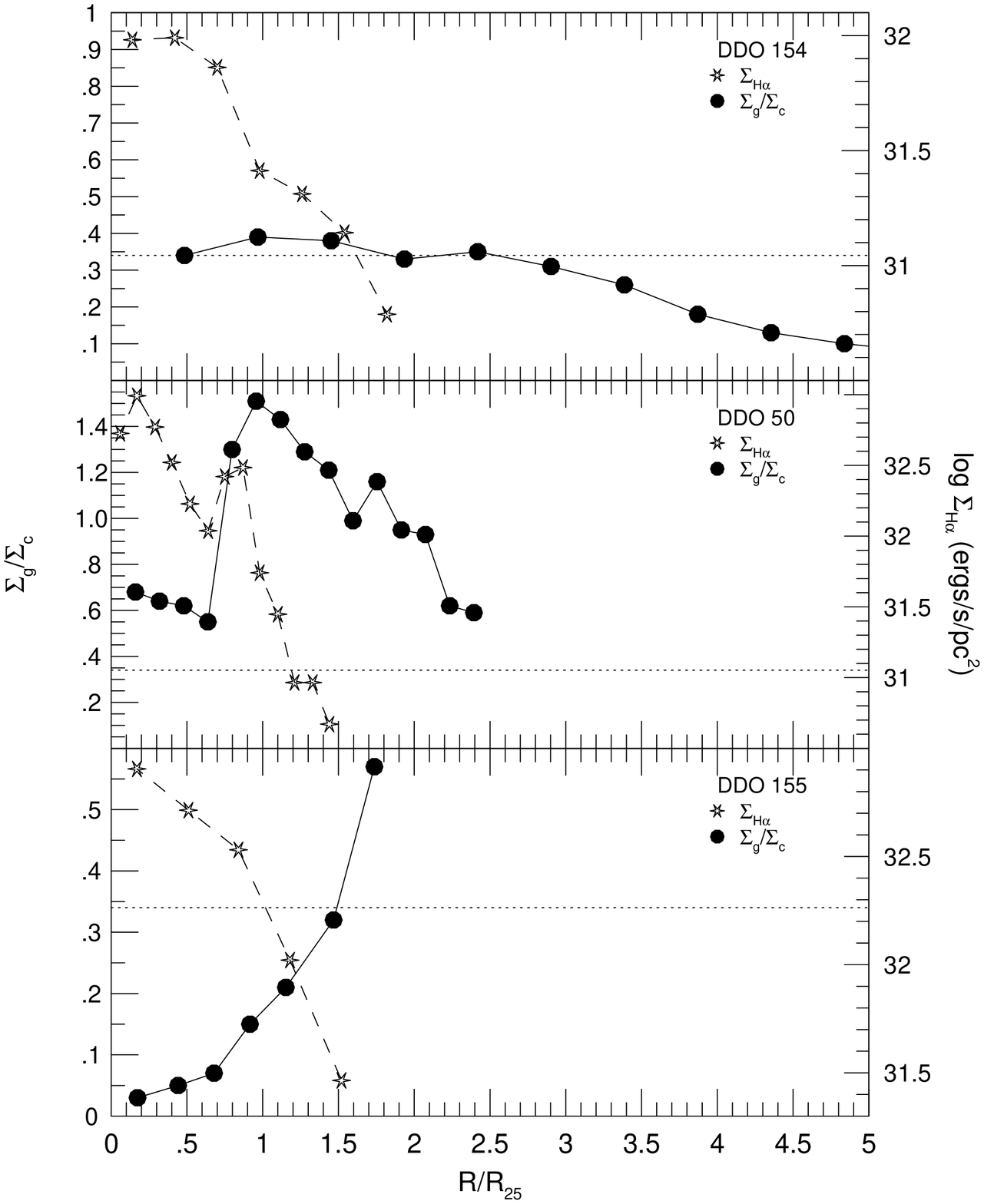}
\caption{\protect\sigrat\ for the thin, rotating gas disk model
and the azimuthally-averaged H$\alpha$ surface
brightness as a function of
radius for the nearby Im galaxy sample.
The radius is normalized to R$_{25}$ for each galaxy.
The horizontal line shows the median value of $\alpha$, \protect\sigrat\
at the largest radius where H$\alpha$ emission is detected, for the
nearby Im galaxy sample.
\label{figim}}
\end{figure}

The ratios of the observed gas densities, \siggas, to the critical gas
densities, \sigcrit, for the single-fluid model are plotted versus
radius as solid lines in Figure \ref{figim} for the nearby Im galaxy
sample.  Also plotted for comparison are the \ha\ surface brightnesses
which are proportional to the star formation rates (Hunter \& Gallagher
1986).  The horizontal dotted line in each panel is the median value of
$\alpha$ determined for our sample of Im galaxies.  This figure can be
compared directly to Kennicutt's Figure 11 for Sc spirals.
 
One can see that with the exception of DDO 50, the ratio \sigrat\ is low
throughout these galaxies compared to normal Sc spirals.  Like Sextans
A, \sigrat\ is generally less than 0.7 in most irregulars.  By contrast,
\sigrat\ is often greater than 1 in the interior regions of Sc spirals,
even reaching $\sim3$ in some cases; it drops to 0.7 and lower only
beyond the radius where star formation is detected.
 
In 6 of the 7 nearby irregulars, \sigrat\ is low in the centers of the
galaxies, higher throughout most of the optical disks, and then lower
again in the outer regions.  The exception is DDO 168.  The degree to
which the center is depressed, and the amount and distance over which
the rise and fall occur, vary from galaxy to galaxy.  In DDO 154, for
example, the central depression is very small and it is followed by a
general and gradual fall off.  Carignan \& Beaulieu (1989) interpreted
the HI distribution in this galaxy as consistent with a low level star
formation rate, although the rate relative to the galaxy's optical size
is in fact at the high end of the range for normal irregulars (Hunter
1997).  Part of the apparent smoothness of \sigrat\ in DDO 154 could be
from resolution affects since there are only 4 HI beamwidths within
\rhii.
 
At the other extreme, \sigrat\ in DDO 50 rises by a factor of $\sim$2.4
at a radius of 2 kpc and then drops off rapidly after that.
Central depressions in the radial distribution of \sigrat\ were
also seen in a high percentage of the irregular galaxy sample
examined by van Zee \et\ (1997). Roughly
half of Kennicutt's Sc spiral galaxies follow this general pattern; the
other half have high \sigrat\ in their centers and the ratio declines
with radius.  Curiously, the surface brightness in \ha\ is highest in
the centers of the irregulars where there is a dip in the \sigrat\
ratio.  In Sextans A the \ha\ emission in the center of the galaxy is
indeed primarily diffuse emission, but in other systems true HII regions
are found in the center. However, in some cases perhaps
the high levels of ionization in the centers of these
galaxies deplete the 
atomic gas there, 
in which case the central dip in \sigrat\ is not related to 
the state of the gas before the star formation began. 
 
The median value of $\alpha$, which is \sigrat\ at \rhii, is
0.34
for the
nearby irregular galaxies.
This is significantly lower than 0.7$\pm$0.2 found by
Kennicutt for Sc
spiral galaxies.  If we adopt $\alpha$ as representing the demarcation
between stable and unstable gas in irregulars, then six of the seven
nearby galaxies have star formation taking place where \sigrat\ is
close to or greater than $\alpha=0.3$. The exception is DDO 155 where
\sigrat\ is much lower than this until the outer part of the galaxy, beyond
the point where star
formation is currently taking place.  This general correlation between
star
formation and higher \sigrat\ in most of the galaxies would seem to say
that the thin rotating disk model is successfully predicting a condition for
star formation.  However, if we look further we see that this is not
the case.  In five of the galaxies, including Sextans A, the \sigrat\
distribution implies that star formation should be taking place
further out in radius than it actually is; that is, \sigrat\
exceeds 0.3 even where there is no star formation. Also, star
formation
is occurring at gas densities only marginally
higher, and sometimes lower, than beyond the optical edge.
Thus, there is no one-to-one correspondence between the presence of
star formation and distinctly higher \sigrat.
 
The one exception to the generally low value of \sigrat\ is DDO 50.
In the central 2 kpc of this galaxy, \sigrat\ is comparable to the
peak values in the other irregular galaxies, but beyond 2 kpc,
\sigrat\ jumps to a maximum of 1.5, which is $\sim$3 times higher than
the average peaks in the other galaxies.  Only out at $\sim$7 kpc, or 2.2 R$_{25}$,
does \sigrat\ decline to the lower value that it had in the center.
The star formation activity itself, as traced by \ha\ emission, drops
rapidly where the high value of \sigrat\ begins its gradual decline,
but ends well before \sigrat\ has dropped to a value $\sim$0.3.
Nevertheless, the
current star formation rate in DDO 50 is not extraordinary, nor is the
surface brightness, which would indicate a low past star formation
rate over a long period of time.  In terms of global parameters DDO
50 is a normal irregular.  There is one feature of DDO 50, however,
that is more extreme than in other irregulars. The HI study of DDO 50
by Puche \et\ (1992) revealed numerous large holes in the neutral gas
which were interpreted as resulting from recent energy input by
concentrations of massive stars.  These holes have diameters up to 1.5
kpc.  This should not affect the overall trend of \sigrat, however,
and
the difference in relative gas density does not appear to
have translated into obvious global differences.
 
\subsubsection{Distant Galaxy Sample}
 
\begin{figure}
\vspace{7.0in}
\includegraphics{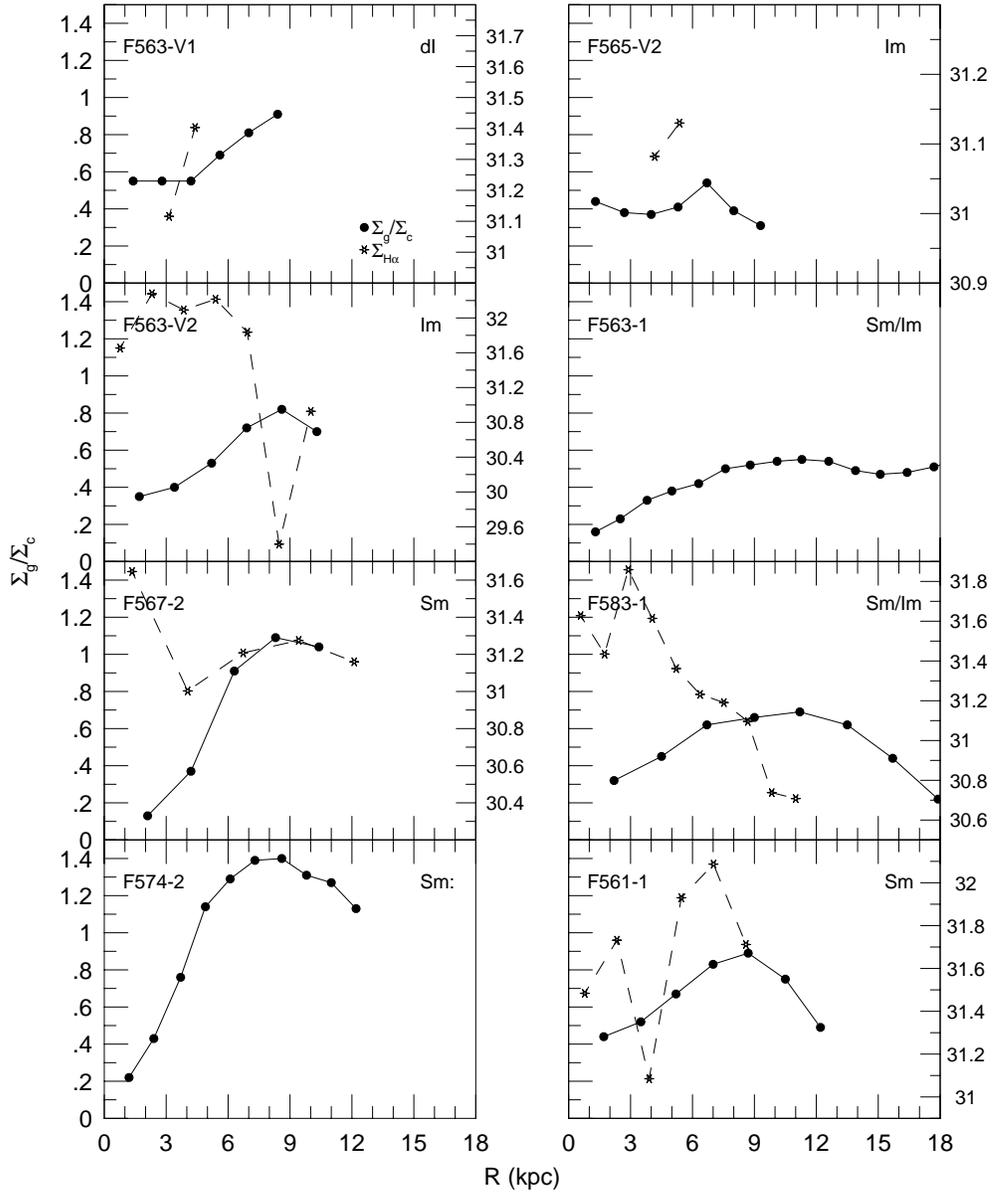}
\caption{\protect\sigrat\ and azimuthally averaged H$\alpha$  surface
brightness as a function of
radius for the distant low surface brightness sample of galaxies.
\label{figf}}
\end{figure}

The relative gas densities and \ha\ radial surface brightnesses
of the distant
galaxy sample are shown in Figure \ref{figf}.
We do not know the detailed molecular content
of these galaxies, but a spectroscopic study of a sample of low surface
brightness galaxies has shown the oxygen abundances in those systems
to be as low as what is usually seen in irregulars
(R\"onnback \& Bergvall 1995, McGaugh 1994).
In addition, as is often the case for irregulars,
CO J$=$1-0 observations failed to detect emission from a sample
of low surface brightness galaxies (Schombert \et\ 1990).
Therefore,
we have made the same assumption for converting
$\Sigma_{HI}$ to $\Sigma_g$ as for the nearby Im sample.
 
We see some similarities and some differences when the distant galaxy
sample is compared with the
nearby Im sample.
First, like the nearby Im sample,
the gas densities in these distant
galaxies are generally low relative to the
critical gas densities in the central parts of the galaxies.
Furthermore,
\sigrat\ rises as one goes out from the center
in all but one system.
However, the central dip in \sigrat\ is generally
more extreme than it is in the nearby Im galaxy sample.
This is also the radial trend seen for low surface brightness
galaxies with ``incipient spiral structure'' examined by
van der Hulst \et\ (1993).
In addition,
the \ha\ surface brightness is high in the center of three of the galaxies,
where the ratio \sigrat\ is low, and it drops with radius.
As for the nearby sample, star formation is not found in
all regions of high \sigrat, nor only where \sigrat\ is highest.
 
However, the most striking difference is that,
unlike most of the nearby Im sample, in 5 of the 8 distant
systems \sigrat\
reaches values $>$0.8, and two reach values $\geq$1.1.
Peak values of $\sim$0.8--1.8 were
found by van der Hulst \et\ (1993) for low surface brightness
spirals (divide $\Sigma_{HI}$ in their
Figure 5 by their $\Sigma_{crit}$ and multiply by 1.52 to convert HI to
total gas using the same assumptions that we used here).
Although \sigrat\ often begins to drop again further out, it does
not drop as rapidly with radius as the star formation activity.
As a consequence, $\alpha$ is higher in these systems: the median
is 0.7, which
is the same as that found for Sc spirals.
Although Kennicutt's Figure 11 is hard to disentangle, it appears
that roughly half of his Sc galaxies have low \sigrat\ in the
center, rising to values of order 1--1.7 and then falling off
with radius beyond that. Thus, these galaxies in our sample,
along with low surface brightness spirals, are behaving more like
normal Sc galaxies than like normal Im galaxies.
 
The natural assumption would be that the distant galaxies are behaving
more like spiral galaxies because they are bigger, both in size
and mass.
However, there are three galaxies in the distant galaxy sample that
have \sigrat\ that are similar to the low values seen in the nearby
Im galaxy sample.
Interestingly, one of these three galaxies (F565-V2)
is also the galaxy with mass model
analysis that shows that it has stellar and dark matter masses
that are in the range of
those determined for Im galaxies (de Blok \& McGaugh 1997).
However, to confound our search for correlations, one of the other galaxies
with low \sigrat\ (F563-1) is also the galaxy of the four with mass models with
the highest stellar and dark matter masses, masses that are significantly
higher than those of normal irregulars.
The third galaxy with low \sigrat\ (F583-1)
has a stellar mass like that of Im systems but a higher dark matter mass.
If one looks at the rotation speed instead of galactic mass,
there is still no correlation to explain why these galaxies
would have lower \sigrat.
 
Thus, the distant galaxy sample has several members that have
gas densities relative to the critical gas densities that are
like those found in many irregular galaxies; \sigrat\ is low in the
center and never rises above the value of $\alpha$ found by
Kennicutt (1989) for spirals. The other members of this sample,
like a sample of low surface brightness spirals,
attain higher values of \sigrat\ and are
similar to some of the high
surface brightness Sc spirals observed by Kennicutt (1989).
What distinguishes the galaxies in this sample
that behave like the irregulars
from those that behave like spirals is not clear.
The defining factor does not seem to be total mass, disk
scale length, rotation speed, or central density.
 
\subsubsection{Global Trends}
 
\begin{figure}
\vspace{7.0in}
\includegraphics{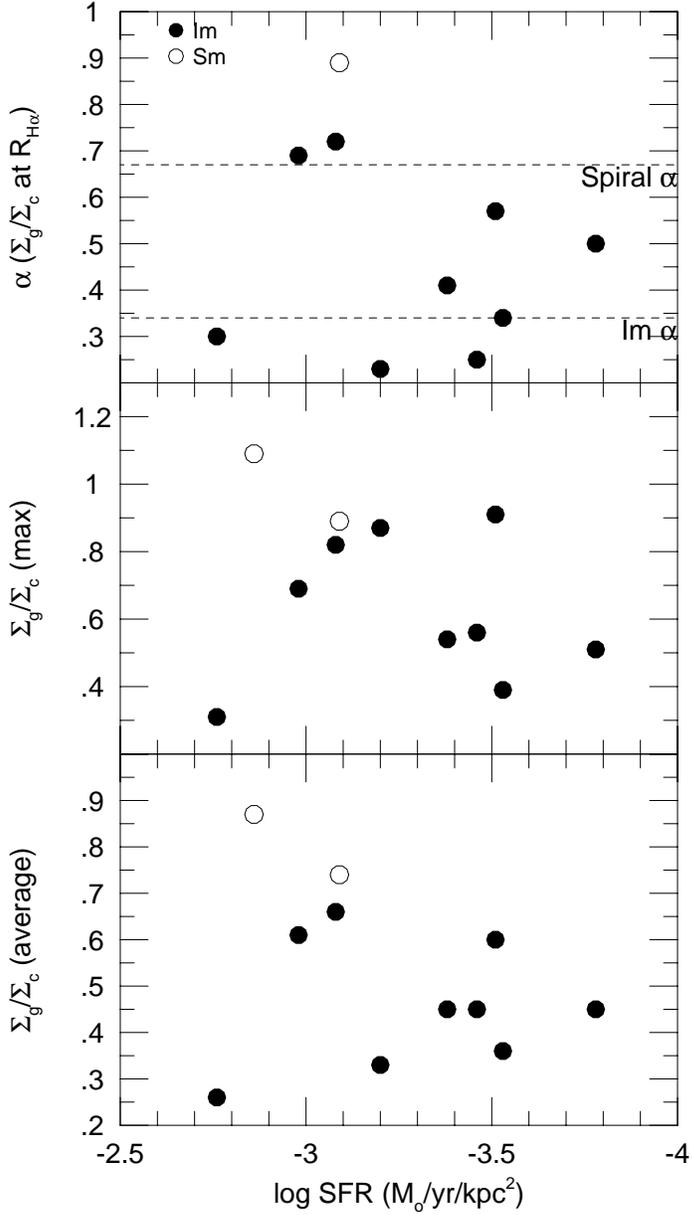}
\caption{Area-weighted average \protect\sigrat\ ratios within
\protect\rhii, maximum \protect\sigrat, and $\alpha$ are plotted as
a function
of the integrated current star formation rate of the galaxies.
The star formation rate is normalized to the \protect\r25\ area
of the galaxy. The top dashed line represents the average Sc spiral
value of $\alpha$ measured by Kennicutt (1989); the bottom dashed line
is the median $\alpha$ found for the nearby sample of irregulars
examined here.
\label{figsfr}}
\end{figure}

In Figure \ref{figsfr} we examine \sigrat\ as a function of the current
level of star formation activity integrated over the galactic disk.  We
might expect that galaxies with higher average \sigrat\ would also have
higher average star formation rates.  However, this does not seem to be
the case.  \sigrat\ is depressed relative to that of Sc spirals, but
within the sample itself we see no correlation with the level of current
star formation activity. Van Zee \et\ (1997) reached a similar
conclusion for their sample.
 
\begin{figure}
\vspace{7.0in}
\includegraphics{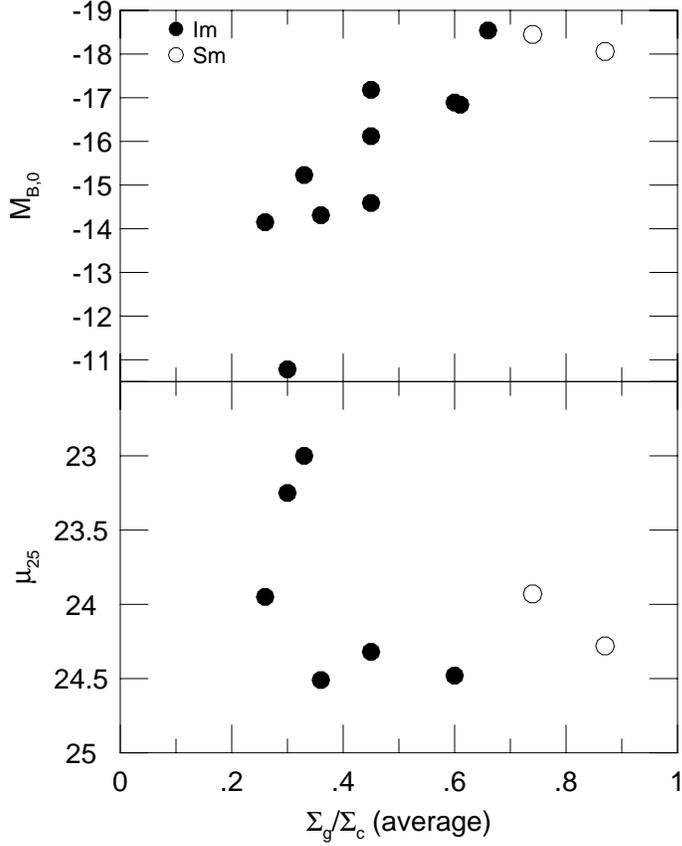}
\caption{Area-weighted average \protect\sigrat\ ratios within
\protect\rhii\ are plotted against
the average B-band surface brightness within \protect\r25\ and
the integrated B-band magnitude of the galaxy.
This surface brightness should be roughly related to the
star formation activity integrated over the past 1 Gyr.
The absolute magnitude of the galaxy has been corrected for extinction.
\label{figsb}}
\end{figure}

Star formation comes and goes within specific regions of a galaxy, and
sometimes the current star formation activity may not represent what has
happened over a more extended period of time.  Therefore, in Figure
\ref{figsb} we plot the \sigrat\ parameter against the optical broad-
band surface brightness, which should depend on the star formation
activity integrated over timescales of 1 Gyr.  Again, we do not see any
correlation. In Figure~\ref{figsb} we also plot the average \sigrat\
against the absolute magnitude of the galaxy. Although the
most luminous few galaxies have the highest average \sigrat, the
rest of the sample has approximately the same \sigrat\ while
covering 7 magnitudes in M$_B$. Thus, M$_B$ is not a significant
predictor of \sigrat, a conclusion that is opposite to that of van Zee \et\
(1997) for their sample of irregulars.
 
\subsection{Thick, Two-fluid Rotating Disk}
 
One limitation to the thin, single-fluid disk model is that it does not
include stars, an obvious component of real galaxies.  A second
limitation is that the model is indeed for a {\it thin} disk.  Irregular
galaxies are known to be thicker than spiral galaxies.  Hodge \&
Hitchcock (1966) measured axial ratios of a large sample of irregular
galaxies and concluded that the ratio of minor to major axes was
0.2--0.4 rather than the 0.2 value usually taken to represent spirals.
Van den Bergh (1988) suggested that irregulars are triaxial with ratios
of 1:0.9:0.4. Puche \et\ (1992) also concluded from an HI study that the
scale height of the neutral gas in DDO 50 is several times greater than
that in spirals.  Thus, irregular galaxies do deviate from the
assumption of a thin disk.
 
\begin{figure}
\vspace{7.0in}
\includegraphics{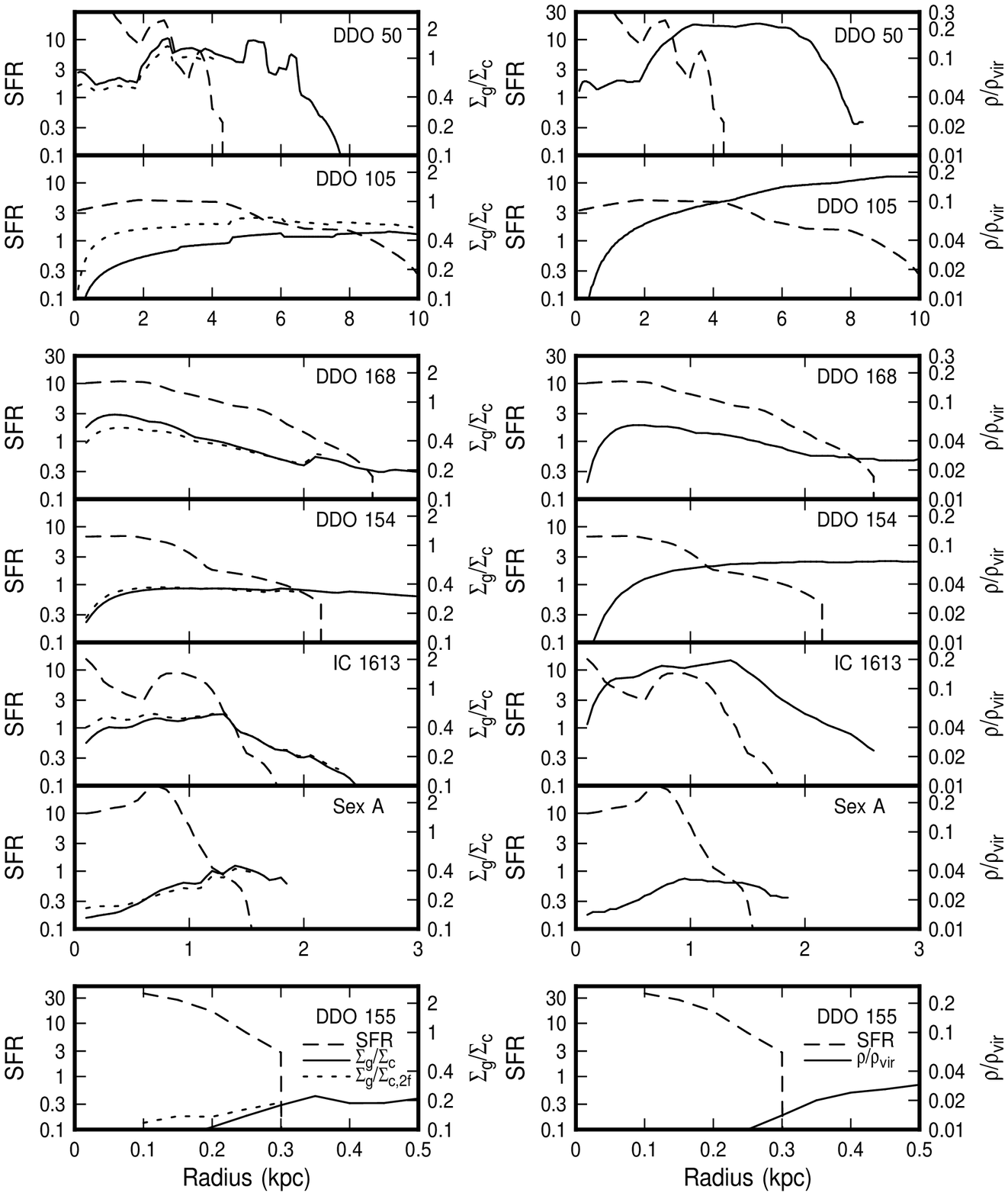}
\caption{In the left panel \sigrat, \sigrateff, and the absolute star formation
rate (SFR) in M\solar yr$^{-1}$
are plotted as a function of radius for the nearby irregular galaxy
sample.
In the right panel we plot the SFR and $\rho/\rho_{vir}$.
\sigcrit\ is the critical column density for a thin rotating gas disk;
\sigeff, the same for a thick, two-fluid rotating disk.
$\rho_{vir}$ is the critical virial density.
All values along the y-axis are plotted on a log scale.
\label{figbe1}}
\end{figure}

We consider now two-fluid instabilities in a thick disk,
and this is applied
to the nearby Im galaxy sample, for which a thick disk should be
more appropriate.  The short-dashed lines on the left
in Figure \ref{figbe1} give \sigrateff\ for this model, taken equal to
the inverse of $Q_{eff}$ in Elmegreen (1995b).  The solid line is
\sigrat\ from before.  We assume gaseous and stellar scale heights of
200 pc and 400 pc, and use the stellar surface density from the B-band
brightness along with $M/L_V$ from model galaxies with monotonic star
formation rates (Larson \& Tinsley 1978).  The results are very similar
to the conventional \sigrat\ because the stellar surface densities are
very low in these galaxies; indeed, they are nearly pure gas systems in
terms of mass (compare $\Sigma_g$ and $\Sigma_s$ in Figure \ref{figbe2}
and see the mass models for DDO 154 in Carignan \& Beaulieu 1989).
In this case, the small addition to self-gravity from the
stars almost exactly compensates for the small loss of
self-gravity from the large thickness.
This
is very different from the situation for giant spiral galaxies, whose
disks are relatively thin and dominated by stars.
In giant spirals,
\sigrateff\ is typically larger
than \sigrat\ by a factor of $\sim2$, making the difference between
spirals and irregulars even larger.

\subsection{Summary of the Rotating Disk Models}
 
We find that the gas densities in irregular galaxies are
low relative to the critical density necessary for the gas to become
unstable within the context of the rotating disk models.  This is true
whether one considers only gas or gas plus stars.  Even the maximum
\sigrat\ in most irregulars is lower than the ratio at the star
formation limit in Sc galaxies.  In addition, the median ratio of
\sigrat\ at the radius of the last detectable \ha\ emission in these
galaxies is lower than that found for spiral galaxies by about a factor
of two.  \sigrat\ in the star forming regions are not much higher than
\sigrat\ near the edges, although star formation is still found where
\sigrat\ is highest.  The exception to the latter result is in the
centers of the galaxies, where \sigrat\ is usually depressed but the
\ha\ surface brightness is often high.  More puzzling is that in many
cases star formation ends in the outer regions before \sigrat\ or
\sigrateff\ drop significantly compared to their values in the interior
regions where star formation is taking place.
 
Physically we are seeing \sigrat\ rise with \siggas\ in the region where
the rotation curves are rising because then $\kappa$ and $\Sigma_c$ are
constant.  Where the rotation curves flatten out, \sigrat\ varies as the
radius times the gas surface density because $\kappa\propto1/R$ so
$\Sigma_c \propto 1/R$.  Values of \sigrat\ are lower than the cutoff
ratio for star formation in spirals found by Kennicutt (1989) because
the gas surface density in irregulars is lower at small radii.  However,
in the outer parts of the irregulars, \sigrat\ remains high as the star
formation rate drops because the gas column density does not drop as
quickly as the star formation.  This is not the case in spiral galaxies,
and so this situation provides a unique test of the various models.
 
Our sample of distant, low surface brightness galaxies provided a
conflicting comparison. Most of these galaxies are larger systems
than those in our Im sample, and they had values of \sigrat\ that
in some cases reached as high as those in some
normal, high surface brightness Sc galaxies. However, several other
systems in this sample had \sigrat\ that were similar to those in
the Im sample. What distinguished these galaxies from the rest
of the sample is not clear but does not appear to be the total
dark matter mass, rotation speed, or disk scale length.
 
We explored the relationship between an area-weighted average gas
density relative to the critical density and the integrated star
formation rate per unit area in our sample of galaxies.  The expectation
was that a galaxy having more gas at higher densities relative to the
critical density would be able to form stars at a higher rate.  However,
we found no such correlation.  This is exemplified by DDO 50 in which
the gas density is significantly higher relative to the predicted
critical gas density from the other irregulars, but the current star
formation rate and optical surface brightness in DDO 50 are not
unusually high.
 
Although the gravitational instability model qualitatively fits the
radial trends of star formation in irregular galaxies in some
ways, the model fails in many details.  Kennicutt (1989) found the same
for Sc galaxies.  For irregulars, the model fails to predict where star
formation ends and which galaxies have higher overall star formation
rates.  For spirals, the model sometimes fails at small radii.  For
example, in M33 and NGC 2403 the observed gas density lies below
\sigcrit\ for much of the inner star-forming disk (Kennicutt 1989,
Thornley \& Wilson 1995).
 
The sample of galaxies discussed here included only those that have
measurable rotation velocities.  There are some galaxies like Leo A in
which rotation is almost non-existent and the gas motion is primarily
random (Young \& Lo 1996).  This was also true of the outer part of DDO
155 (Carignan \et\ 1990).  In these systems the model of a rotating disk
is not applicable at all.  Here it would make more sense to discuss
stability in terms of the virial density, or to use the criterion for the
existence of a cool thermal phase.  We address these models next.
 
\section{Virial and Thermal Models}
\label{sec:thermal}
 
Another measure of gaseous self-gravity in a galaxy comes from the ratio
of the gas volume density to the total virial density from matter in all
forms, including stars and dark matter.  This virial density comes from
the rotation curve and is taken here equal to
$\rho_{vir}=(15/4\pi)\left(V/R\right)^2/G$ for a spherical distribution
of matter (mostly in a dark halo); $V$ is the local rotation speed and
$R$ is the radius in the galaxy.  Elmegreen et al.\ (1996) discussed the
merits of considering $\rho_{vir}$ as a contributor to the Jeans length
for star formation in irregular galaxies.  It may be more useful than
either \sigcrit\ or $\Sigma_{c,2f}$ for irregulars because these
galaxies are thick, often in solid body rotation, and their dynamics is
dominated by dark matter (Carignan \& Beaulieu 1989).
Phenomenologically, one expects $\rho/\rho_{vir}$ to be relatively large
in regions of rapid star formation because then self-gravity can
overcome any background galactic tidal or shear forces.  For a thin
disk, this ratio becomes similar to \sigrat.  Note that the absolute
value of $\rho/\rho_{vir}$ in a star-forming region has not been
calibrated from observations, so only the relative distribution of this
ratio is considered here.
 
The results of applying the models to the nearby Im galaxy sample
are shown on the right in Figure \ref{figbe1}, along with
the star formation rate in M\solar\ yr$^{-1}$.  The trends of
$\rho/\rho_{vir}$ with radius are similar to the trends of \sigrat\
and \sigrateff, so $\rho/\rho_{vir}$ has the same ambiguous
connection to a star formation threshold.
 
These ambiguities lead us to consider another criterion for the onset
of star formation, which is the critical combination of parameters
that allows the interstellar medium to contain cool diffuse clouds
(Elmegreen \& Parravano 1994). This model postulates that too low a
thermal gas pressure, or too high a gaseous heat source, will permit
only the warm (several $\times 10^3$ K) phase of HI to exist.  Then
the condensation of this gas to cool HI clouds and ultimately cold
H$_2$ clouds becomes unlikely unless violent processes increase the
pressure locally (like sequentially triggered star formation or
end-of-bar gas flows). Criteria for this pure-warm state were
evaluated in Elmegreen \& Parravano (1994) and a simple approximation
to the result was given in Elmegreen (1997).  It leads to a critical
(lower limit) to the gas column density for the existence of cool gas:
\begin{equation}
\Sigma_{th}=7\left({{\Sigma_g}\over{\theta\Sigma_{total}}}
\right)^{0.5}\left(I/I_\odot\right)^{0.6} \;\;{\rm M}_\odot \;
{\rm pc}^{-2}
\end{equation}
Here $I/I_\odot$ is the ratio of the average intensity in the midplane
of the irregular to the average intensity in the midplane near the Sun;
$\Sigma_{total}$ is the sum of the column densities of the gas and
stars inside the gas layer (this quantity appears in the expression
for the total midplane gas pressure), and $\theta$ is the ratio of the
thermal pressure to the total gas pressure.  To get the stellar
surface density inside the gas layer, we used the total stellar
surface density, $\Sigma_s$, determined from the surface brightness as
described above, and multiplied this by the ratio of the gas to the
stellar velocity dispersions (see derivation in Elmegreen 1989).  The
velocity dispersion of the stars was taken to be twice that for the
gas, so the total surface density in the gas layer was in fact
$\Sigma_{total}=\Sigma_g +0.5\Sigma_s$.  We also assume $\theta=0.1$,
which is approximately the value for the solar neighborhood; $\theta$
is unknown for irregular galaxies.
 
The left-hand side of Figure \ref{figbe2} plots
$\Sigma_g/\Sigma_{th}$
versus radius for the nearby sample.  $\Sigma_{th}$ depends on $I$,
the average stellar radiation field as a function of position in the
midplane of each galaxy.  To get this quantity, we first determined
the radiation field inside a model of the Milky Way disk, including
extinction, as discussed in Elmegreen \& Parravano (1994).  This gave
the ratio of the average intensity in the midplane at the solar radius
to the intensity viewed perpendicular to the plane from outside the
Milky Way.  With this ratio we calibrated the midplane average
intensity using the effective surface brightness at the Sun's radius.
The V surface brightness of the solar neighborhood was taken to be
22.98 magnitudes arcsec$^{-2}$ as viewed from outside in the direction
of the pole (Allen 1963).  Then we calculated the average-midplane and
perpendicular intensities in an irregular galaxy.  We assumed the local
volume emissivity was proportional to the observed surface brightness
at each radius, $M_V$, and integrated over all angles at each radius
in the midplane to get the relative average intensity and the
perpendicular intensity, as observed from infinity outside the disk.
The midplane intensity was calibrated to absolute units by determining
the ratio of the average midplane intensity to the perpendicular
intensity.  With the observed perpendicular intensity (i.e., the
observed surface brightness, corrected for inclination) of the galaxy
and the perpendicular intensity of the Sun's neighborhood, we get the
ratio of the average midplane intensity in the irregular to that
near the Sun.  The integration over all angles includes azimuthal and
polar angles.  For each point on the line of sight, we used the
midplane density obtained from the surface density divided by twice
the scale height, and modified by a Gaussian in z.  This density
appears in the local absorption coefficient.  We used the stellar
density on the line of sight for this integration from the volume
emissivity modified by a Gaussian in z too, using the stellar scale
height.
 
For the Milky Way we used a gas scale height of 0.1 kpc and a stellar
scale height of 0.2 kpc.  The disk scale length is 4 kpc with a
maximum disk extent of 24 kpc; the Sun is at a radius of 8.5 kpc.  The
absorption was taken to be 1.5 optical depths per kpc for a unit gas
density.  For the irregulars, we assumed gas and stellar scale heights
that were twice those for the Milky Way and absorption per unit gas
that was 0.3 times that for the Milky Way to account for the lower
metal and dust content of irregulars.  The gas density was determined
from the observed column density divided by twice the scale height
of the gas.
 
In most of the galaxies, the ratio \siggas/$\Sigma_{th}$ is $\leq$1,
suggesting that the gas prefers to be in the warm phase over much of
the disk.  This is a consequence of the low total surface density,
since the pressure scales with the total surface density times the gas
surface density.  Since the stellar surface density is small compared
to that of the gas, the total surface density is essentially the gas
surface density.  Thus, the pressure is approximately $P=33.5 k_B
\Sigma_g^2$ for \siggas\ in units of M\solar pc$^{-2}$, where $k_B$ is
Boltzman's constant.  The total pressure in the solar neighborhood is
3$\times 10^4 k_B$ and the equivalent \siggas\ to give this is
30 M\solar pc$^{-2}$.  Thus, $P \approx P\solar (\Sigma_g/30 M\solar
pc^{-2})^2 $.  Since \siggas\ in the irregulars, shown in
Figure~\ref{figbe2}, is usually much less than 30 M\solar pc$^{-2}$,
the average midplane pressures are very low in the irregular galaxies, as
low as 1\% of the solar neighborhood pressure in some cases.  This
means that it is hard to sustain a cool phase in the HI gas,
even in the presence of relatively low radiation fields.
Note that these pressures are independent of any assumptions about
magnetic field strengths or supernova rates, as long as the gas
layer is in hydrostatic equilibrium.
 
Without a significant amount of dark matter in the disks of irregular
galaxies, the small value of the ratio $\Sigma_g/\Sigma_{th}$ implies
that most of the HI is warm, as mentioned above.  This result is
consistent with the findings of Young \& Lo (1996).  In Leo A
($=$DDO69) they found two HI components, one with a high velocity
dispersion (9 \kms) which they identified as a warm gas component, and
one with a low velocity dispersion (3.5 \kms) which they identified as
a cold gas component.  The cold gas component was associated with
current star formation while most of the HI gas was found in the warmer
gas component.
 
There is generally no correspondence between the azimuthally-averaged
star formation rate and $\Sigma_g/\Sigma_{th}$ as a function of
radius.  In DDO 50, DDO 154, and Sextans A, $\Sigma_g/\Sigma_{th}$
rises above a value of 1 where the star formation is dropping. In DDO
168 the star formation drops off even though $\Sigma_g/\Sigma_{th}$
remains around 1.  Thus the star formation mechanism is not very
dependent on the mean thermal properties of the HI.
The presence of significant dark matter in the disk can affect this
conclusion, so we discuss the potential role of dark matter
in Section \ref{sect:dm}.
 
\section{A Cloud-Growth Criterion Based on Shear}
 
The critical column density used by Kennicutt (1989)
was based on the stability of thin disks for which there is
constant competition between self-gravity, pressure, and
Coriolis forces (which come from angular momentum conservation).
These three forces introduce the three physical
variables in the $Q$ parameter ($\equiv\Sigma_c/\Sigma_g$),
namely, $\Sigma_g$, $c$, and
$\kappa$, respectively.  Other physical processes of cloud
formation can depend on slightly different combinations of the
physical variables.
 
Consider a situation where
clouds are constantly
trying to form because of self-gravity and
energy dissipation, and the growth occurs with
streaming
motions along interstellar
magnetic field lines. Then the
Coriolis force can be somewhat less important
than in the above analysis
because magnetic fields easily
transfer angular momentum away from the cloud (Elmegreen 1987, 1991).
What is more important in this case
is the time available for the cloud to grow in the presence of shear.
 
Shearing perturbations grow at the rate $\pi G\Sigma_g/c$, and this
growth is most effective between the times $-1/A$ and $1/A$ for
Oort constant $A=0.5Rd\Omega/dR$, which is the local shear rate.
Thus the total amplitude of the growth from some initial
perturbation $\delta \Sigma_0$ is
\begin{equation}
\delta \Sigma_{peak}\sim\delta\Sigma_0 \exp\left({{2\pi G\Sigma_g}
\over{cA}}\right).
\end{equation}
For the instability to be significant, a perturbation must grow
by a sufficiently large factor, say $\sim100$. Thus
\begin{equation}
\exp\left({{2\pi G\Sigma_g}\over{cA}}\right)>100
\end{equation}
or
\begin{equation}
{{\Sigma_g}\over{\Sigma_{c,A}}} >1
\end{equation}
for critical column density
\begin{equation}
\Sigma_{c,A}={{2.5Ac}\over{\pi G}}
\label{eq:sa}
\end{equation}
based on the Oort $A$ constant instead of $\kappa$.
 
This change from $\kappa$ to $A$ makes very little difference in the
critical column density for a flat rotation curve.
If we write the rotation curve as $V\propto R^\alpha$,
then
\begin{equation}
{{\Sigma_{c,A}}\over{\Sigma_{c}}}={{0.88(1-\alpha)}\over
{(1+\alpha)^{1/2}}}.
\end{equation}
For a flat rotation curve such as is usually seen in spiral galaxies,
$\alpha=0$, and the two thresholds
are the same to within 12\%.  For a slowly rising rotation
curve or any region of low shear,
$\alpha$ becomes small and
$\Sigma_{c,A}$ can be much less than $\Sigma_c$.
For example, in M33 the rotation curve is slowly rising, giving
$\alpha\sim0.3$ (Newton 1980). Then
$\Sigma_{c,A}=0.54\Sigma_c$, which is in excellent agreement
with $\Sigma_g$ observed by Wilson et al. (1991)
even though this is the region where \sigrat\ fails to
predict the observed high star formation rate.
This derivation of $\Sigma_{c,A}$ and the motivation for it
originally appeared in Elmegreen (1993b).
 
This process seems
particularly well suited to irregular galaxies because of their
slowly rising rotation curves.
Figure \ref{fig:dm} (left) shows values of the growth parameter
$\Sigma_g/\Sigma_{c,A}$ derived according to equation \ref{eq:sa}.
In all cases,
$\Sigma_g/\Sigma_{c,A}$ is much closer to 1 than
$\Sigma_g/\Sigma_{c}$ was.  There is still not a perfect
agreement between the growth parameter and the star formation
rate, however. In particular $\Sigma_g/\Sigma_{c,A}$ remains
high even after star formation ends. But, the overall level of
$\Sigma_g/\Sigma_{c,A}$ in the regions of star formation is better than for
any other criterion given in this paper.
This suggests that cloud formation
in irregular and other galaxies
may involve more of a competition between self-gravity and shear
than between self-gravity and Coriolis forces.
 
\section{The Role of Dark Matter}
\label{sect:dm}
 
Many irregular galaxies have a higher dark matter fraction than
spiral galaxies overall, by a factor of 10 or more, and some may have
larger dark matter central densities too (Kormendy 1988, Carignan \&
Freeman 1988).  If the dark matter fraction is relatively high in the
disks of irregular galaxies, then dark matter could affect star
formation and our estimate of $\alpha$.
 
There are several observations of dynamical masses for the galaxies
studied here.  In DDO154, Carignan \& Freeman (1988) and Carignan \&
Beaulieu (1989) found that dark matter is 90\% of the total mass at the
last measured point of the HI rotation curve, which is about four times the
size of the optical disk.  They also found $M/L_B\sim1$ for the stellar
disk and obtained a central dark matter density of $\rho_{DM}\sim0.02$
M$_\odot$ pc$^{-3}$.  In IC1613, Lake \& Skillman (1989) found
$M/L_B\sim0.5$ and $\rho_{DM}\sim0.001$ M$_\odot$ pc$^{-3}$ in the inner
regions.  In DDO 155, Carignan, Beaulieu, \& Freeman (1990) estimated
$M/L_B\sim16$ and $\rho_{DM}\sim0.07$ M$_\odot$ pc$^{-3}$ within 500 pc.
In DDO 168, Broeils (1992) found $M/L_B\sim0.1$ and $\rho_{DM}\sim0.04$
M$_\odot$ pc$^{-3}$ in the luminous disk and $\rho_{DM}\sim0.002$
M$_\odot$ pc$^{-3}$ at $R_{25}$ where $M_{dark}/M_{lum}=3.8$.  He also
estimated in DDO 105 that $M/L_B\sim0.8-5.5$ and $\rho_{DM}\sim0.0008-0.0045$
M$_\odot$ pc$^{-3}$ in the luminous disk and
$M_{dark}/M_{lum}=0.2-2.9$  and
$\rho_{DM}\sim0.0006-0.0013$ M$_\odot$ pc$^{-3}$
at $R_{25}$.
 
Various other results have been obtained for late-type galaxies not included
here.  Van Zee et al (1996) found that $\sim90$\% of the total mass is
dark in UGCA 20.
For NGC 247 (Sd), NGC 300 (Sd), and NGC 3109 (Sm),
Carignan \& Freeman (1985) find, respectively:  $\rho_{DM}\sim$ 0.0033,
0.0042, and 0.0024 M$_\odot$ pc$^{-3}$ in the centers and 0.0026,
0.0029, and 0.0014 M$_\odot$ pc$^{-3}$ at the Holmberg radii.  Also for
these three galaxies, they obtain $M_{halo}/M_{disk}\sim$ 0.85, 0.70,
and 1.52.
However, not all irregular galaxies are dominated by dark matter.
Young \& Lo (1996) found that the dynamical mass
equals the luminous mass in LeoA if it is in equilibrium.
In addition, for the distant
low surface brightness galaxies, de Blok et al. (1996) speculate
and de Blok \& McGaugh (1997) show that those systems are also
dark matter dominated and that
the dark matter is more diffusely distributed than in high surface
brightness galaxies.
 
For a typical dark matter central density of $\rho_{DM}\sim0.003$
M$_\odot$ pc$^{-3}$, the column density of dark matter within $\pm 500$
pc of the midplane is $\sim3$ M$_\odot$ pc$^{-3}$, which is several
times larger than the stellar column densities in many of our galaxies,
and comparable to the gas column densities.  Here we consider how such a
high dark matter fraction in the optical regions of irregular galaxies
can affect disk stability and star formation.
 
Dark matter is likely to be more important for star formation in small
galaxies than in large galaxies for several reasons.  First, small
galaxies have low virial speeds, so the dark matter velocity dispersion
is much smaller than in large galaxies, perhaps within a factor of 2 of
what is expected for the stars in a small galaxy.  Then dark matter
effectively acts like stars in the two-fluid instability, giving extra
self-gravity to small perturbations in the gas.  Second,
the relatively large scale heights of small galaxies implies that
relatively more halo dark matter resides in the disk, again contributing
to disk self-gravity.
 
Third, thick disks with dark matter should have much higher pressures
than thin disk or disks of the same large thickness with only gas and
stars.  The total gas pressure scales directly with the
surface density of matter in all forms within the gas layer.  This
higher pressure may allow the cool HI phase to exist as a precursor to
star formation.  The very low pressure derived in section
\ref{sec:thermal} is suspicious anyway:  considering only the known
stars and gas, the self-gravity of the disk gives a pressure that is
only several per cent of the disk pressure in giant spirals.  The HII
region pressures are probably much larger than this.  A pressure of
$\sim500$k$_B$ for an irregular galaxy (Sect.  \ref{sec:thermal}) gives
an equilibrium HII region density of only 0.024 cm$^{-3}$, considering
that $P(HII)\sim2.1nkT$ for $T\sim10^4$ K. Such an equilibrium HII
region would have an emission measure of only 0.5 cm$^6$ pc even if it
were 1000 pc thick, and then it would be essentially invisible.  In
spiral galaxies, the typical HII regions that are studied
in surveys have a density of $\sim1$ cm$^{-3}$ (as
determined from data in Kennicutt 1988), which places them in rough
pressure equilibrium with the ambient gas (i.e., the HII regions that
stand out in galaxies are the relatively old ones that have expanded to
large sizes, and not the young regions that are still at high pressure).
We might expect the same for HII regions in irregular galaxies, but this
would require a much higher midplane pressure than the self-gravity of
the gas and stars can give alone.  Thus there is probably
more mass in the disks of Im and Sm galaxies than is directly
attributed to HI and stars.
 
The effect of dark matter on star formation may be assessed
by evaluating $\Sigma_{c,2f}$ and $\Sigma_{th}$
for models with higher total disk column densities.
As an example, we assume that the total disk contains 4 times
more mass than the sum of the visible gas and stars, and that
all of this extra mass has the velocity dispersion of the stars,
which is chosen to give the stellar scale height used above,
400 pc.
Figure \ref{fig:dm} (right) shows \sigrateff\
calculated in this manner
for the nearby irregulars, along with
\sigrat\ and the star formation rate from before.
$\Sigma_g/\Sigma_{th}$ with dark matter is not shown
because it increases by exactly a factor of 2 over that in Figure \ref{figbe2},
making $\Sigma_g/\Sigma_{th}$ generally $\sim1$,
since $\Sigma_{th}$ is inversely proportional to the
square root of the
total
disk column density (from the pressure). The change in
\sigrateff\ is more complicated; in many cases the
gas still dominates the instability because of its low
velocity dispersion.
 
Dark matter in the disks of these galaxies clearly makes them
more unstable to star formation. It may not
be a coincidence that the galaxy with evidence for no dark matter, Leo A,
also has a very low star formation rate compared to other irregulars.
But, there is still a problem
with the detailed
distribution of \sigrateff\ relative to the star
formation rate. For example, there are outer disks where star formation
ends but \sigrateff\ remains high, and there are regions where
star formation is prominent but \sigrateff\ is relatively low.
Moreover, the inclusion of dark matter or stars in the
stability criterion for
giant {\it spiral}
galaxies, which is usually based on
\sigrat\ for a pure gas
disk, increases the value of $\alpha$ found at the disk edge
by a factor of $\sim2$, so irregulars would still be more stable than
spiral galaxies by this factor if both contain the expected amounts of total
disk mass.
 
\section{The Relationship of Star Formation to Gas and Stars}
 
\begin{figure}
\vspace{7.0in}
\includegraphics{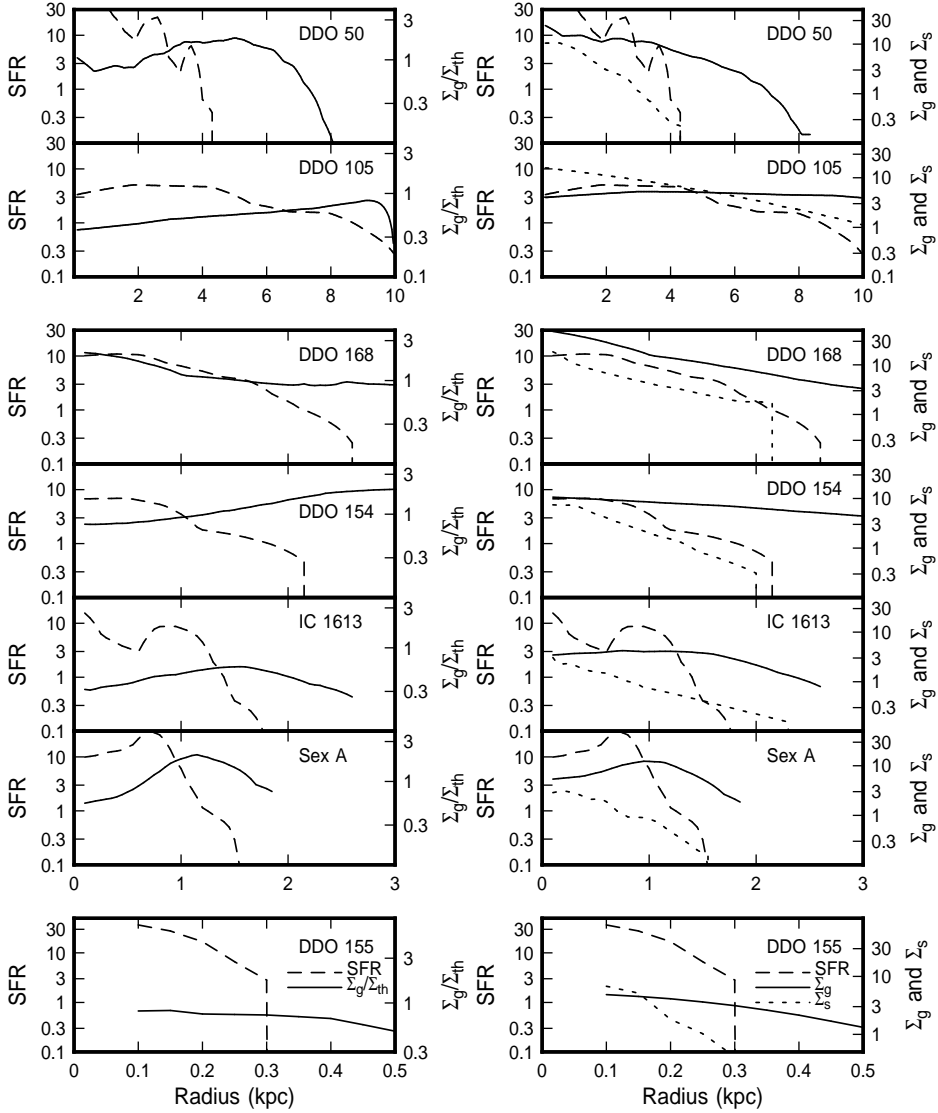}
\caption{On the left is plotted
$\Sigma_g/\Sigma_{th}$ and the star formation rate (SFR)
in M\solar yr$^{-1}$.
On the right, the SFR and observed gas ($\Sigma_g$) and stellar
($\Sigma_s$) surface densities.
$\Sigma_{th}$ is the critical gas column density for the existence
of cool gas in the thermal model.
The densities
are in units of M\protect\solar pc$^{-2}$.
All values along the y-axis are plotted on a log scale.
\label{figbe2}}
\end{figure}

\begin{figure}
\vspace{7.0in}
\includegraphics{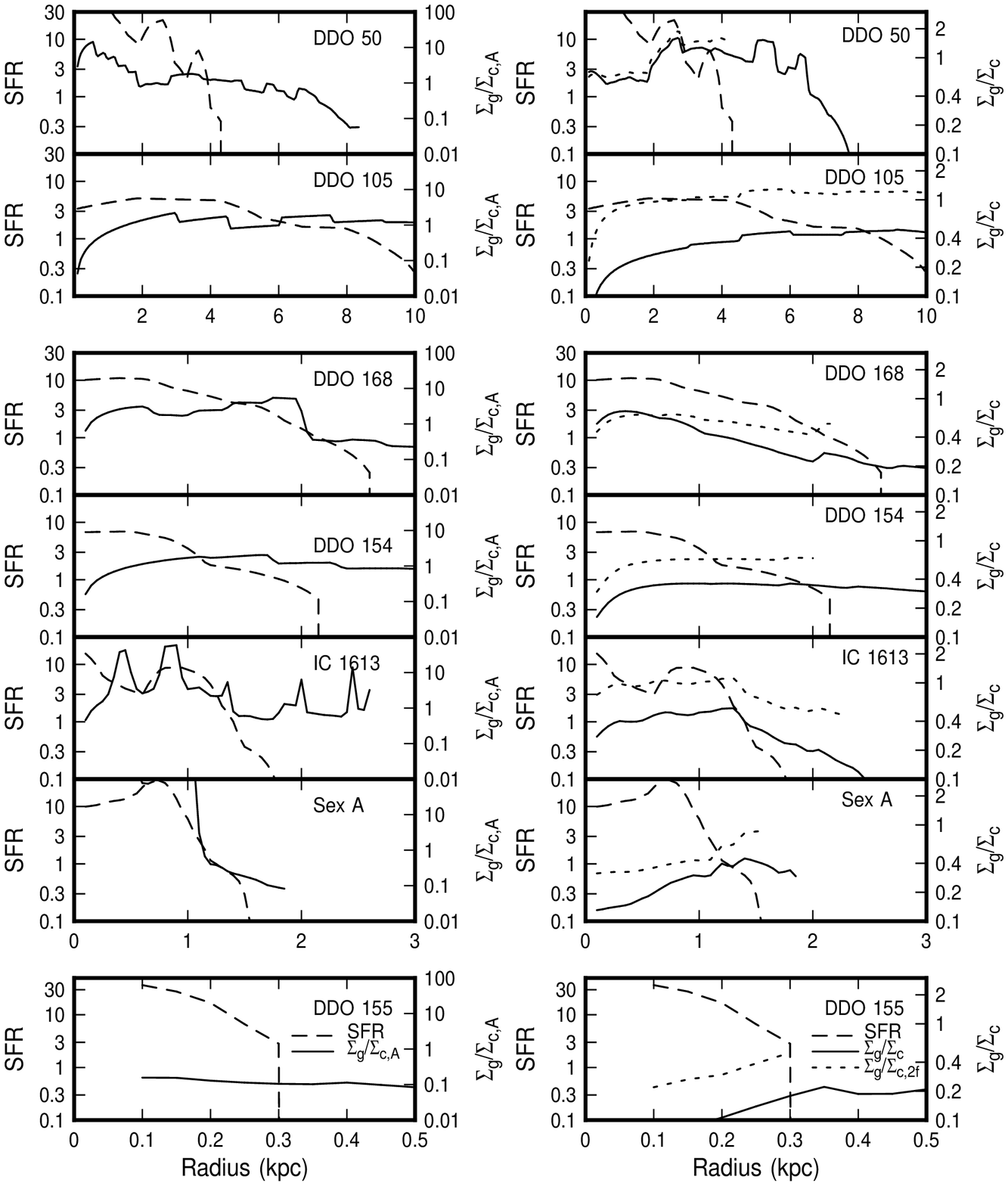}
\caption{Plots on the right show the ratios
of observed gas densities to critical column
densities $\Sigma_{c,2f}$ in the two-fluid
model with dark matter in a thick disk, giving a total disk column density
equal to four times the observed value.  This is compared to
\sigcrit\ for gravitational instabilities in the thin, rotating gas disk
without dark matter.
The plots on the left show the ratio of
observed gas density to critical column density $\Sigma_{c,A}$ for
a cloud-growth criterion based on shear.
The star formation rate (SFR) is also shown.
\label{fig:dm}}
\end{figure}

Since critical densities that depend on local conditions have failed
to predict where star formation occurs and ends, we consider instead simply
the observed gas and stellar densities, which we plot in
Figure~\ref{figbe2} along with the star formation rate.
Ryder \& Dopita (1994) suggested that star formation is closely
related to the mean HI surface density in the disks of spiral galaxies.
However, we see no correlation between the star formation activity
and \siggas\ as a function of radius in irregulars.
 
Perhaps instead of a varying critical density,
we are dealing with a constant threshold column density that does not
depend on \sigrat, but is universal (see references in section 1).
Then the star formation rate
simply occurs where the gas density is greater than this threshold.
We can see a hint of this from Figure~\ref{figbe2}
where the star formation rate is plotted along with \siggas.
In all of these galaxies, the place where the star formation rate
has dropped to about 0.3 M\solar yr$^{-1}$ also has \siggas\ equal
to
approximately 3 M\solar pc$^{-2}$. Could this apply to spiral galaxies
as well? We can use Kennicutt's value of $\alpha\sim0.7$ for Sc
galaxies
at the radius beyond which star formation is not
detected. Then \sigcrit $= 0.7 c \kappa/\pi G$. If we take c$=$7 \kms\
and $\kappa = \sqrt{2} (220)$ \kms\ R$^{-1}$, then
\sigcrit $= 0.0236 /R(kpc)$ in gm cm$^{-2}$ which is
112 M\solar pc$^{-2} / R(kpc)$.
For $R(kpc)\sim20$ where star formation in spiral galaxies
ends, this value of
\sigcrit\ is only slightly higher than what we find for the
irregulars,
and in fact it is higher by about the same factor that $\alpha$
is lower in the irregulars. This implies
that a star formation model with a simple threshold gas column density
may
have some merit.
 
The {\it best} correlation between the azimuthally-averaged
current star formation activity and any other galaxy property
discussed here is with the total stellar density $\Sigma_s$, measured from the
surface brightness.  This is shown on the right in Figure
\ref{figbe2}, where the star formation rate (dashed line) and
$\Sigma_s$ (dotted line)
generally lie close to each other throughout the whole galaxy.
But, is this correlation a cause or an effect? If the star formation
rate is roughly
constant with radius over timescales of a Gyr for whatever physical
reason, we would expect to
see such a correlation.
However, the correlation could also imply a physical connection---a
feedback mechanism between the
older stars and the star formation process.
Perhaps star formation needs turbulence to make clouds, and the stars
increase the turbulence through gravitational stirring, stellar winds,
and radiation pressure.
 
The relationship between star formation activity and $\Sigma_s$
is similar to what is seen in spiral galaxies.
Ryder \& Dopita (1994) observed that star formation closely follows
the old stellar
mass surface density in spirals although they saw less of a correlation in V-band
than in I-band. Also in spirals,
the molecular gas, as traced by CO, follows the blue starlight
(Young \& Scoville 1982).
 
If star-induced star formation plays a larger role in facilitating cloud
formation in irregulars than in other galaxies, then the lack of stars when the
irregular first formed
would result in a system that was slower to begin forming stars.
Observationally, star formation rates in early-type spirals have
been gradually dropping over the Hubble time (Schmidt 1959; Sandage 1986;
Larson 1992; Kennicutt, Tamblyn, \& Congdon 1994), while those in
late-type spirals and irregulars have remained roughly constant
(see, for example, Kennicutt \et\ 1994;
Tosi \et\ 1991; Greggio \et\ 1993; Marconi \et\ 1995; Tolstoy 1996).
For those irregulars that
were isolated, the lack of external stimulations would allow
a sustained low level of star formation, as Bothun \et\ (1993) have
suggested, as long as the star-induced star formation process was
not so efficient that the feedback from stars
was able to run away (Larson 1996).
 
An alternative explanation for the correlation between
star formation rate and old star density is that
star formation always gives an exponential disk,
regardless of the current star formation rate or the value of an instability
criterion. This could result from viscous evolution, for example,
with a star formation rate proportional to the viscosity
(Lin \& Pringle 1987; Yoshii \& Sommer-Larsen 1989; Saio \& Yoshii 1990).
Then the
correlation between \sigrat\ and star formation rate for spiral galaxies
would be coincidental and not necessarily the same for irregulars.
 
For example,
$\Sigma_g/\Sigma_c\sim1$
may be controlled by spiral-wave induced disk accretion in giant
galaxies, a process that would not work well in irregulars because of their
relatively large scale heights. Spiral waves
prefer thin galaxies where the velocity dispersion is low relative to the
rotation speed. These waves, along with viscosity,
produce torques that drive gas accretion,
always keeping $\Sigma_c$ at the threshold of spiral-forming instabilities.
If
$\Sigma_g/\Sigma_c\sim1$ because of spiral-driven accretion, then this ratio
could be anything in irregular galaxies and it would have little
direct connection to star formation.
For an exponential gas disk with a flat rotation curve
and nearly constant velocity dispersion,
\sigrat$\propto \Sigma_g/\kappa\propto
Re^{-R/R_s}$ for disk scale length $R_s$. This function is small at
small radii, rises to a peak at $R_s$, and then decreases in
the outer part of the galaxy, as observed for \sigrat. The
observation
that \sigrat\ is greater than 0.7 or some other constant inside \rhii\
would follow entirely from the fact that
$\Sigma_g/\Sigma_c\sim1$ generally.
In this interpretation, \sigrat\ is more related to the formation of
spiral arms than to star formation activity.
The results presented here for irregular galaxies
support this interpretation because they show no strong correlations
between \sigrat\ and star formation activity once spiral arms are
absent.
 
\section{Summary}
 
Irregular galaxies are an interesting and unique test of models for the physical laws
regulating the formation of gas clouds that can form stars in galaxies.
Irregulars lack the spiral density waves and tidal shears that
contribute
to gravitationally induced instabilities.
Irregulars, unlike spirals, also have azimuthally-averaged
gas surface densities that do not drop off as fast as star formation.
 
We have calculated several model critical gas densities
for instabilities that would lead naturally to cloud and
star formation
in irregular galaxies.
These models considered thin pure-gas disks, thick gas+star disks,
three dimensional systems including dark matter, the
thermal properties of the gas, and shear-regulated cloud formation.
The resulting critical densities for cloud formation were compared
with the observed gas densities in a sample of irregular galaxies.
The ratio of observed to critical gas density as a function of radius
was also compared with
the azimuthally-averaged current star formation rate as traced
by \ha\ emission and with the broad-band stellar surface brightness,
a tracer of star formation integrated over a timescale of $\sim$ 1 Gyr.
 
We found that the ratios of observed to critical gas densities are
lower in irregulars than in spiral galaxies by a factor of $\sim2$
in all cases but the shear-regulated model.
This suggests that the gas in irregulars is closer to
stability, even though star formation is occurring.
However, no model was able to predict with any accuracy where
the
star formation actually occurs, and especially where it
ends.
These two items suggest that other processes are
important for cloud and star formation.
An obvious mechanism is sequentially triggered star formation
driven by the mechanical energy
input from concentrations of massive stars.
Applications of this mechanism to small
galaxies have been discussed extensively (e.g., Gerola,
Seiden \& Schulman 1980; Athanassoula 1994).
Another process
is random gas compression from turbulence. Both of these
processes have a sensitivity to the critical density like the
gravitational instability model, but do not involve large-scale
instabilities directly.  They could in fact dominate the star
formation process when spiral arms are not present.
 
Kennicutt (1989) and Thornley \& Wilson (1995) have pointed out
that the gravitational instability model for the thin, pure-gas disk
also fails for several late-type spirals. In particular the gas
in the inner parts of M33 and NGC 2403 is below \sigcrit\ and yet
star formation is taking place there. Thornley \& Wilson also
suggest that other processes, such as stochastic star formation,
may be responsible for initiating cloud formation in these
regions. However, both of these galaxies also have weak spiral
density waves, so it may be the spiral density waves that are
playing the key role here.
 
We also considered the possibility that
dark matter contributes a higher fraction of the total disk mass in
irregulars than in spirals, perhaps by a factor of two.  This
enhances disk instability in the two-fluid model and allows the
formation of cool HI,
a necessary precursor to star formation.  A high dark matter fraction
in the disk
also seems reasonable for irregulars because the ratio of dark
matter to total matter is larger in most of these galaxies than in giant
spirals.
However, if we include dark matter in irregulars, then we should
also include stars and other disk mass in the instability threshold for giant spirals.
When this is done, there is still a factor of two discrepancy between $\alpha$
for spirals and irregulars.
 
The possibility that star formation requires a ${\it constant}$
threshold gas column density rather than a value that scales with the
epicyclic frequency, as in the usual analysis, was also considered. A
threshold of $\sim3$ M$_\odot$ does indeed seem to predict where
star formation stops in the outer disks of both irregular galaxies and
giant spirals, and it may also apply to low surface brightness disks
that have gas column densities this low or lower throughout.
The physical reason behind a constant density threshold is
not known, but such a criterion was also proposed in the model of
stochastic self-propagating star formation by Seiden \& Schulman
(1990).
 
The best correlation we found was between
the current star formation activity
and the stellar surface density.
This could
indicate a feedback mechanism in which
energy input by stars to the interstellar medium facilitates cloud
formation for future generations of stars. If this feedback plays a key
role in the formation of clouds, then this could also
explain why irregular
galaxies
were slow to begin forming stars after they first coalesced out
of the primordial gas.
 
Another possibility is that the stellar surface density correlation
results from a star formation process that always gives an exponential
disk.  Then $\Sigma_g/\Sigma_c$ may be $\sim1$ because of processes
related to spiral arm generation rather than star formation.  In this
case, the correlation between \sigrat\ and star formation in giant
galaxies is not causal, and a different \sigrat\ in non-spiral galaxies,
such as irregulars, is possible.  A good test of this explanation may
come from flocculent galaxies or the low surface brightness
samples of large galaxies, which have only weak spiral arms.  So
far, flocculent galaxies and a majority of the large, low
surface brightness systems seem to have $\alpha\sim0.7$, the same as giant
spirals.  This comes from Kennicutt's (1989) study, which included
several flocculents, van der Hulst et al.'s (1993) study of
low surface brightness spirals, and our study of large, low surface
brightness galaxies without obvious spiral arms.
However, if the low $\alpha$ that we have found for
irregulars is the result of star formation not strongly related to
\sigrat, and if the real feedback process giving constant \sigrat\ for
giant galaxies is related to spiral arm generation, and only indirectly
related to star formation, then flocculent and large low surface
brightness galaxies could lack a
detailed correlation between \sigrat\ and star formation just like the
irregulars.
 
The comparisons with models that we have made here have all dealt
with azimuthally-averaged quantities.
However, in irregular galaxies, neither the star formation nor the HI
is uniformly
distributed across the galaxy. Thus, azimuthal-averages of HI and
\ha\ are only rough approximations to radial trends (see, for example,
van Zee \et\ 1997). This may be particulary important in galaxies
with low shear, such as irregulars. We have suggested that 
sequentially triggered star formation driven by mechanical energy
input from concentrations of massive stars is likely to be an important
mechanism in cloud formation in these small galaxies.
A consequence of the low levels of shear is that the holes formed
by these massive stars or by any other mechanism
will last a longer time and smoothing of the
gas will take longer than in spirals.
It is the gas density of clumps in the ISM relative to the critical gas
density that is important, not the average gas density, in determining
the instability of that gas to cloud formation.
This implies that
if the surface filling factor of dense gas is about 0.5 in irregulars,
then the
effective value of $\alpha$ from the instability analysis
should be increased by 1/0.5 to account for instabilities in only
the dense gas. This makes $\alpha$ where the gas is
more like the value
for giant spiral galaxies, but it
also illustrates how the condensations
cannot come only from instabilities in an initially
uniform medium, because then $\alpha$ would be too low again.
We will explore
this idea in more detail in a future paper.

\acknowledgments
 
Part of this work was done while ALB was a participant in the 1996
MIT Astronomy Field Camp for Undergraduates held at Lowell Observatory,
and part of this work formed ALB's MIT Bachelor of Science Thesis.
DAH would like to thank the Instituto de Astrof\'isica de Canarias
for hospitality while part of this paper was being written.
 
\clearpage
 
\begin{deluxetable}{lrrrrrrr}
\scriptsize
\tablecaption{Properties of the Sample Galaxies \label{tabgal}}
\tablewidth{0pt}
\tablehead{
\colhead{Galaxy} & \colhead{Type\tablenotemark{a}}
& \colhead{D\tablenotemark{b}}
& \colhead{E(B$-$V)$_f$\tablenotemark{c}}
& \colhead{E(B$-$V)$_T$\tablenotemark{c}}
& \colhead{M$_{B_T}$\tablenotemark{d}}
& \colhead{$\mu_{25}$\tablenotemark{e}}
& \colhead{R$_{25}$\tablenotemark{f}} \nl
\colhead{} & \colhead{}
& \colhead{(Mpc)}
& \colhead{} & \colhead{} & \colhead{}
& \colhead{(mag arcsec$^{-2}$)} & \colhead{(\arcsec)}
}
\startdata
IC1613	& IB(s)m  & 0.73 & 0.005    & 0.06\xx & $-$14.6 & 24.3\xxxx & 284 \nl
DDO 50	& Im	  & 3.2  & 0.02\xx  & 0.07\xx & $-$16.7 & 23.8\xxxx & 188 \nl
DDO 105 & IB(s)m: & 15.6 & 0\xx     & 0.05\xx & $-$17.2 & 25.1\xxxx & 71 \nl
DDO 154 & IBsm	  & 4.0  & 0.008\xx & 0.06\xx & $-$14.3 & 24.5\xxxx & 62 \nl
DDO 155 & Im	  & 1.1  & 0.01\xx  & 0.06\xx & $-$10.8 & 23.3\xxxx & 38 \nl
DDO 168 & IBm	  & 3.5  & 0\xx     & 0.05\xx & $-$15.2 & 23.0\xxxx & 144 \nl
Sextans A & IB(s)m & 1.3 & 0.02\xx  & 0.02\xx & $-$14.2 & 24.0\xxxx & 144 \nl
F561-1	& Sm	  &  72  & 0.04\xx  & 0.09\xx & $-$18.8 & 23.9\xxxx & 21 \nl
F563-1	& Sm/Im   &  52  & 0.02\xx  & 0.07\xx & $-$18.3 & 24.5\xxxx & 23 \nl
F563-V1 & dI	  &  58  & 0.03\xx  & 0.08\xx & $-$17.2 & 24.5\xxxx & 13 \nl
F563-V2 & Irr  & 71  & 0.04\xx	    & 0.09\xx & $-$18.9 & \nodata\xxxx & 23 \nl
F565-V2 & Im	  &  55  & 0.02\xx  & 0.07\xx & $-$16.4 & 25.2\xxxx & 10 \nl
F567-2	& Sm	  &  86  & 0.03\xx  & 0.08\xx & $-$18.4 & 24.3\xxxx & 9 \nl
F574-2	& Sm:	  & 101  & 0.01\xx  & 0.06\xx & $-$18.7 & 24.5\xxxx & 9 \nl
F583-1	& Sm/Irr  &  37  & 0.03\xx  & 0.08\xx & $-$17.2 & \nodata\xxxx & 21 \nl
\enddata
\tablenotetext{a}{Galactic morphological classes are from RC3
and Schombert \et\ (1992).}
\tablenotetext{b}{Based on an H$_0$ of 65 \kms\ Mpc$^{-1}$ where
V$_{GSR}$ are obtained from the HI references or RC3.
The exceptions are DDO 155
and IC1613
where the distances are taken from de Vaucouleurs \& Moss (1983)
and McAlary, Madore, \& Davis (1984).}
\tablenotetext{c}{E(B$-$V)$_f$ is
foreground reddening from Burstein \& Heiles (1978, 1984).
Total reddening E(B$-$V)$_T$ is
foreground plus internal where the internal reddening
is taken to be 0.05 for cases where an observational estimate does
not exist.}
\tablenotetext{d}{Total absolute B magnitude corrected for extinction.
Magnitudes come from RC3 and de Blok \et\ (1995, 1996).}
\tablenotetext{e}{Average B surface brightness within the 25
magnitude arcsec$^{-2}$ surface brightness level, corrected for
extinction. Magnitudes come from RC3 and de Blok \et\ (1995, 1996).}
\tablenotetext{f}{Radius of the galaxy measured to the 25 B-magnitude
arcsec$^{-2}$, in arcsec, corrected for reddening.
Values are determined from surface photometry given by
Carignan \et\ (1990), Carignan \& Beaulieu (1989), Broeils (1992),
and images obtained at Lowell Observatory.
For the distant sample values are taken from
de Blok \et\ (1995, 1996) and not corrected for reddening.}
\end{deluxetable}
 
\clearpage
 
\begin{deluxetable}{lrrrrr}
\scriptsize
\tablecaption{Adopted Parameters and References \label{tabref}}
\tablewidth{0pt}
\tablehead{
\colhead{Galaxy}
& \colhead{pa\tablenotemark{a}} & \colhead{i\tablenotemark{a}}
& \colhead{$c$\tablenotemark{b}}
& \colhead{HI Reference} & \colhead{$\mu$ Reference} \nl
\colhead{}
& \colhead{(\arcdeg)} & \colhead{(\arcdeg)}
& \colhead{(km/s)}
& \colhead{} & \colhead{}
}
\startdata
IC1613
&  58 & 38 & 7.5\xx &
Lake \& Skillman 1989\tablenotemark{c} & Ables 1971 \nl
DDO 50
&  30 & 40 & 6.8\xx &
Puche \et\ 1992 & This paper \nl
DDO 105
& 250 & 61 & (9)\xx &
Broeils 1992 & This paper \nl
DDO 154
& 38 & 57 & (9)\xx &
Carignan \& Beaulieu 1989 & Carignan \& Beaulieu 1989 \nl
DDO 155
&  47 & 47 & 9.5\xx &
Carignan \et\ 1990 & Carignan \et\ 1990 \nl
DDO 168
& 328 & 76 & (9)\xx &
Broeils 1992 & Broeils 1992 \nl
Sextans A
& 50 & 36 & 9.0\xx &
Skillman \et\ 1988 & Hunter \& Plummer 1996 \nl
F561-1
&  55 & 24 & (9)\xx &
de Blok \et\ 1996 & \nl
F563-1
& $-$19 & 25 & (9)\xx &
de Blok \et\ 1996 & \nl
F563-V1
& $-$40 & 60 & (9)\xx &
de Blok \et\ 1996 & \nl
F563-V2
& $-$32 & 39 & (9)\xx &
de Blok \et\ 1996 & \nl
F565-V2
& $-$155 & 60 & (9)\xx &
de Blok \et\ 1996 & \nl
F567-2
& 119 & 20 & (9)\xx &
de Blok \et\ 1996 & \nl
F574-2
&  53 & 30 & (9)\xx &
de Blok \et\ 1996 & \nl
F583-1
& $-$5 & 63 & (9)\xx &
de Blok \et\ 1996 & \nl
\enddata
\tablenotetext{a}{Position angles and inclinations are taken from the
HI references and are used in doing elliptical photometry of \ha\ images.
The exception is DDO 50 where the HI position angle did not make sense
for the optical image; instead we used a position angle determined
from the broad-band optical image.}
\tablenotetext{b}{Velocity dispersion of the gas. Values in parenthesis
are assumed; values for DDO 155 and DDO 50 are galaxy-wide averages.}
\tablenotetext{c}{Values in Table IV of Lake \& Skillman (1989) are
labeled as Jy/beam \kms\ but appear to actually be 10$\times$mJy/beam \kms.}
\end{deluxetable}

\clearpage
 
\begin{deluxetable}{lrrrrrr}
\footnotesize
\tablecaption{Star Formation Rates and Gas Densities \label{tabgas}}
\tablewidth{0pt}
\tablehead{
\colhead{Galaxy} & \colhead{log SFR/area\tablenotemark{a}}
& \colhead{\protect\rhii\tablenotemark{b}}
& \colhead{${R_{H\alpha}}\over{FWHM_{beam}}$\tablenotemark{c}}
& \colhead{$\langle$\sigrat$\rangle$\tablenotemark{d}}
& \colhead{(\protect\sigrat)$_{max}$}
& \colhead{$\alpha$\tablenotemark{e}} \nl
\colhead{} & \colhead{(M\solar/yr/kpc$^{-2}$)}
& \colhead{(kpc)}
& \colhead{}
& \colhead{}
& \colhead{}
& \colhead{}
}
\startdata
IC1613	  & $-$3.46\xxxx &  1.8\x & 14.2\xxxx & 0.45\xx & 0.56\xxxx & 0.25 \nl
DDO 50	  & $-$2.88\xxxx &  4.3\x & 69.8\xxxx & 1.14\xx & 1.51\xxxx & 1.14 \nl
DDO 105   & $-$3.78\xxxx & 10.4\x & 10.5\xxxx & 0.45\xx & 0.51\xxxx & 0.50 \nl
DDO 154   & $-$3.53\xxxx &  2.2\x &  3.3\xxxx & 0.36\xx & 0.39\xxxx & 0.34 \nl
DDO 155   & $-$2.24\xxxx &  0.3\x &  3.9\xxxx & 0.30\xx & 0.57\xxxx & 0.38 \nl
DDO 168   & $-$3.20\xxxx &  2.6\x & 11.0\xxxx & 0.33\xx & 0.87\xxxx & 0.23 \nl
Sextans A & $-$2.76\xxxx &  1.2\x &  4.2\xxxx & 0.26\xx & 0.31\xxxx & 0.30 \nl
F561-1	  & $-$3.09\xxxx &  8.6\x &  1.9\xxxx & 0.74\xx & 0.89\xxxx & 0.89 \nl
F563-1	  & \nodata\xxxx & \nodata & \nodata\xxx & \nodata\xx & 0.56\xxxx & \nodata \nl
F563-V1   & $-$3.51\xxxx &  4.4\x &  1.2\xxxx & 0.60\xx & 0.91\xxxx & 0.57 \nl
F563-V2   & $-$3.08\xxxx & 10.0\x &  2.2\xxxx & 0.66\xx & 0.82\xxxx & 0.72 \nl
F565-V2   & $-$3.38\xxxx &  5.4\x &  1.5\xxxx & 0.45\xx & 0.54\xxxx & 0.41 \nl
F567-2	  & $-$2.86\xxxx & 12.1\x &  2.2\xxxx & \nodata\xx & 1.09\xxxx & \nodata \nl
F574-2	  & \nodata\xxxx & \nodata & \nodata\xxx & \nodata\xx & 1.40\xxxx & \nodata \nl
F583-1	  & $-$2.98\xxxx & 10.9\x &  4.7\xxxx & 0.61\xx & 0.69\xxxx & 0.69 \nl
\enddata
\tablenotetext{a}{Star formation rate per unit area where the
area is that within R$_{25}$. The star formation rate is determined
from L$_{H\alpha}$ according to the formula in Hunter \& Gallagher (1986).}
\tablenotetext{b}{The radius, deprojected to the plane of the galaxy,
at which the furthest H$\alpha$ emission is detected in our
\ha\ images.}
\tablenotetext{c}{The shorter axis of the synthesized beam was used.}
\tablenotetext{d}{The area-weighted average of \sigrat\ for the region
of the galaxy interior to \rhii.}
\tablenotetext{e}{$\alpha$ is \sigrat\ at \rhii\ (after Kennicutt
1989).
For F567-2, the HI map did not go out as far as \rhii, so a value of $\alpha$
could not be determined there.}
\end{deluxetable}

\clearpage
 
\clearpage
 
\begin{references}
 
\reference{} Ables, H.\ 1971, PUSNO, 2d Ser., 20, Part 4
\reference{} Allen, C.\ W.\ 1963, Astrophysical Quantities (London:
Oxford Univ.\ Press), 269
\reference{} Athanassoula, E.\ 1984, Phys.Reps., 114, 319
\reference{} Athanassoula, E.\ 1994,
in Dwarf Galaxies, edited by G.\ Meylan and P.\ Prugniel
(Garching bei M\"unchen,ESO), p 525
\reference{} Binggeli, B.\ 1994,
in Dwarf Galaxies, edited by G.\ Meylan and P.\ Prugniel
(Garching bei M\"unchen,ESO), p 13
\reference{} Bothun, G.\ D., Schombert, J.\ M., Impey, C.\ D., Sprayberry,
D., McGaugh, S.\ S.\ 1993, AJ, 106, 530
\reference{} Boulanger, F. \& Viallefond, F. 1992, A\&A, 266, 37
\reference{} Broeils, A.\ 1992, Ph.D.\ thesis, Rijksuniversiteit Groningen
\reference{} Buat, V., Deharveng, J.\ M., \& Donas, J.\ 1989, A\&A, 223, 42
\reference{} Burstein, D., \& Heiles, C.\ 1978, ApJ, 225, 40
\reference{} Burstein, D., \& Heiles, C.\ 1984, ApJS, 54, 81
\reference{} Byrd, G., Rautiainen, P., Salo, H., Buta, R., \&
Crocker, D.\ A.\ 1994, AJ, 108, 476
\reference{} Caldwell, N., Kennicutt, R., Phillips, A.C. \&
Schommer, R.A. 1992, ApJ, 370, 526
\reference{} Carignan, C., \& Beaulieu, S.\ 1989, ApJ, 347, 760
\reference{} Carignan, C., Beaulieu, S., \& Freeman, K.\ C.\ 1990, AJ, 99, 178
\reference{} Carignan, C., \& Freeman, K.\ C.\ 1985, ApJ, 294, 494
\reference{} Carignan, C., \& Freeman, K.\ C.\ 1988, ApJ, 332, L33
\reference{} Cohen, R., Dame, T., Garay, G., Montani, J., Rubio, M., \&
Thaddeus, P.\ 1988, ApJ, 331, L95
\reference{} de Blok, W.\ J.\ G., \& McGaugh, S.\ S.\ 1997, MNRAS, in press
\reference{} de Blok, W.\ J.\ G., McGaugh, S.\ S., \& van der Hulst,
J.\ M.\ 1996, MNRAS, 283, 18
\reference{} de Blok, W.\ J.\ G., van der Hulst, J.\ M., \& Bothun,
G.\ D.\ 1995, MNRAS, 274, 235
\reference{} de Vaucouleurs, G., de Vaucouleurs, A., Corwin, H., Buta,
R., Paturel, G., \& Fouqu\'e, P.\ 1991, Third Reference Catalogue of
Bright Galaxies (New York:Springer) (RC3)
\reference{} de Vaucouleurs, G., \& Moss, C.\ 1983, ApJ, 271, 123
\reference{} Dopita, M.\ A., Mathewson, D.\ S., \& Ford, V.\ L.\ 1985, ApJ, 297, 599
\reference{} Elmegreen, B.\ G.\ 1982, ApJ, 253, 634
\reference{} Elmegreen, B.\ G.\ 1987, ApJ, 312, 626
\reference{} Elmegreen, B.\ G.\ 1989, ApJ, 338, 178
\reference{} Elmegreen, B.\ G.\ 1991, ApJ, 378, 139
\reference{} Elmegreen, B.\ G.\ 1992, in Star Formation in Stellar 
Systems, edited
by G.\ Tenorio-Tagle, M.\ Prieto, and F.\ S\'anchez,
(Cambridge Univ. Press,Cambridge), p 381
\reference{} Elmegreen, B.\ G.\ 1993a, ApJ, 419, L29
\reference{} Elmegreen, B.\ G.\ 1993b, in
Star Forming Galaxies and their Interstellar Media,
ed. J. Franco, F. Ferrini, and G. Tenorio-Tagle, Cambridge
University Press, p. 337
\reference{} Elmegreen, B.\ G.\ 1994a, ApJ, 433, 39
\reference{} Elmegreen, B.\ G.\ 1994b, ApJ, 427, 384
\reference{} Elmegreen, B.\ G.\ 1994c, ApJ, 425, L73
\reference{} Elmegreen, B.\ G.\ 1995a, in
The 7th Guo Shoujing Summer
School on Astrophysics: Molecular Clouds and Star Formation,
ed. C. Yuan and Hunhan You, (Singapore: World Press) p 149
\reference{} Elmegreen, B.\ G.\ 1995b, MNRAS, 275, 944
\reference{} Elmegreen, B.\ G.\ 1997, in Starburst Activity in
Galaxies, editted by J.\ Franco, R.\ Terlevich, \&
A.\ Serrano, Revista
Mexicana de Astron. and Astrofis. (Serie Conferencias), 6, 165
\reference{} Elmegreen, B.\ G., \& Parravano, A.\ 1994, ApJ, 435, L121
\reference{} Elmegreen, B.\ G., Elmegreen, D.\ M., Salzer, J.\ J.,
\& Mann, H.\ 1996, ApJ, 467, 579
\reference{} Fall, S.M., \& Efstathiou, G. 1980, MNRAS, 193, 189
\reference{} Gammie, C.\ F.\ 1996, ApJ, 462, 725
\reference{} Gallagher, J.\ S., \& Hunter, D.\ A.\ 1984, ARA\&A, 22, 37
\reference{} Gallagher, J.\ S., Hunter, D.\ A., \& Tutukov, A.\ V.\
1984, ApJ, 284, 544
\reference{} Gerola, H., Seiden, P.\ E., \& Schulman, L.\ S.\ 1980,
ApJ, 242, 517
\reference{} Goldreich, P., \& Lynden-Bell, D.\ 1965, MNRAS, 130, 97
\reference{} Greggio, L., Marconi, G., Tosi, M., \& Focardi, P.\ 1993,
AJ, 105, 894
\reference{} Guiderdoni, B.\ 1987, A\&A, 172, 27
\reference{} Hodge, P.\ W.\ 1971, ARA\&A, 9, 35
\reference{} Hodge, P.\ W.\ 1978, ApJS, 37, 145
\reference{} Hodge, P.\ W., \& Hitchcock, J.\ L.\ 1966, PASP, 78, 79
\reference{} Hodge, P.\ W., \& Kennicutt, R.\ C.\ 1983, ApJ, 267, 563
\reference{} Hodge, P.\ W., Lee, M.\ G., \& Gurwell, M.\ 1990, PASP,
102, 1245
\reference{} Hunter, D.\ A.\ 1997, PASP, in press
\reference{} Hunter, D.\ A., \& Gallagher, J.\ S.\ 1985, ApJS, 58, 533
\reference{} Hunter, D.\ A., \& Gallagher, J.\ S.\ 1986, PASP, 98, 5
\reference{} Hunter, D.\ A., Gallagher, J.\ S., \& Rautenkranz, D.\ 1982,
ApJS, 49, 53
\reference{} Hunter, D.\ A., Hawley, W.\ N., \& Gallagher, J.\ S.\ 1993, AJ,
106, 1797
\reference{} Hunter, D.\ A., \& Plummer, J.\ D.\ 1996, ApJ, 462, 732
\reference{} Jog, C., \& Solomon, P.M. 1984, ApJ, 276, 114
\reference{} Kennicutt, R.\ C.\ 1988, ApJ, 334, 144
\reference{} Kennicutt, R.\ C.\ 1989, ApJ, 344, 685
\reference{} Kennicutt, R.\ C., Tamblyn, P., \& Congdon, C.\ W.\ 1994,
ApJ, 435, 22
\reference{} Kormendy, J.\ 1988, in Origin, Structure and Evolution of
Galaxies, ed. Fang Li Zhi (Singapore: World Scientific), p. 252
\reference{} Kroupa, P. 1995, MNRAS, 277, 1491
\reference{} Lake, G., \& Skillman, E.\ D.\ 1989, AJ, 98, 1274
\reference{} Larson, R.\ B.\ 1992, in III Canary Islands Winter School
of Astrophysics: Star Formation in Stellar Systems,
edited by G.\ Tenorio-Tagle, M.\ Prieto, and F.\ S\'anchez
(Cambridge:Cambridge University Press), p 125
\reference{} Larson, R.\ B.\ 1996, in The Interplay Between Massive Star
Formation, the ISM, and Galaxy Evolution, edited by D.\ Kunth, B.\
Guiderdoni, M.\ Heydari-Malayeri, and T.\ X.\ Thuan
(Gif-sur-Yvette, France:Editions Frontieres), p 3
\reference{} Larson, R.\ B., \& Tinsley, B.\ M.\ 1978, ApJ, 219, 46
\reference{} Lin, D.N.C., \& Pringle, J.E. 1987, MNRAS, 225, 607
\reference{} Madore, B.\ F., van den Bergh, S., \& Rogstad, D.\ H.\
1974, ApJ, 191, 317
\reference{} Marconi, G., Tosi, M., Greggio, L., \& Focardi, P.\ 1995, AJ, 109, 173
\reference{} McAlary, C.\ W., Madore, B.\ F., \& Davis, L.\ E.\ 1984,
ApJ, 276, 487
\reference{} McGaugh, S.\ S.\ 1994, ApJ, 426, 135
\reference{} McGaugh, S.\ S., Schombert, J.\ M., \& Bothun, G.\ D.\
AJ, 109, 2019
\reference{} Mouschovias, T.\ Ch.\ 1981, in Fundamental Problems in
Theory of Stellar Evolution, ed.\ D.\ Q.\ Lamb \& D.\ N.\ Schramm
(Dordrecht:Reidel), p 27
\reference{} Newton, K.\ 1980, MNRAS, 190, 689
\reference{} Parker, E.\ N.\  1966, ApJ, 145, 811
\reference{} Pildis, R.\ A., Schombert, J.\ M., \& Eder, J.\ A.\
1997, ApJ, 481, 157
\reference{} Price, J.\ S., Mason, S.\ F., \& Gullixson, C.\ A.\
1990, AJ, 100, 420
\reference{} Puche, D., Westpfahl, Brinks, E., \& Roy, J-R.\ 1992, AJ, 103,
1841
\reference{} Quirk, W.\ J.\ 1972, ApJ, 176, L9
\reference{} Reaves, G.\ 1952, PhD thesis, University of California
\reference{} Reaves, G.\ 1997, private communication
\reference{} Romeo, A.\ B.\ 1992, MNRAS, 256, 307
\reference{} R\"onnback, J., \& Bergvall, N.\ 1995, A\&A, 302, 353
\reference{} Rubio, M., Garay, G., Montani, J., \& Thaddeus, P.\
1991, ApJ, 368, 173
\reference{} Ryder, S.\ D., \& Dopita, M.\ A.\ 1994, ApJ, 430, 142
\reference{} Safronov, V.\ S.\ 1960, Ann.d'Ap., 23, 979
\reference{} Saio, H., \& Yoshii, Y. 1990, ApJ, 363, 40
\reference{} Salpeter, E.\ 1955, ApJ, 121, 161
\reference{} Sandage, A.\ 1971, ApJ, 166, 13
\reference{} Sandage, A.\ 1986, A\&A, 161, 89
\reference{} Schild, R.\ 1977, AJ, 82, 337
\reference{} Schmidt, M.\ 1959, ApJ, 129, 243
\reference{} Schombert, J.\ M., \& Bothun, G.\ D.\ 1988, AJ, 95, 1389
\reference{} Schombert, J.\ M., Bothun, G.\ D., Impey, C.\ D.,
\& Mundy, L.\ G.\ 1990, AJ, 100, 1523
\reference{} Schombert, J.\ M., Bothun, G.\ D., Schneider, S.\ E.,
\& McGaugh, S.\ S.\ 1992, AJ, 103, 1107
\reference{} Schombert, J.\ M., Pildis, R.\ A., Eder, J.\ A.\ \& Oemler,
A.\ 1995, AJ, 110, 2067
\reference{} Seiden, P.\ E.\  1983, ApJ, 266, 555
\reference{} Seiden, P.\ E.\ , \& Schulman, L.\ S.\ 1990, Adv. Phys., 39, 1
\reference{} Shlosman, I., \& Begelman, M.C. 1989, ApJ, 341, 685
\reference{} Skillman, E.\ D.\ 1987, in Star Formation in Galaxies,
ed.\ C.\ Persson (Washington:GPO), p 263
\reference{} Tammann, G.\ A.\ 1980, in Dwarf Galaxies, edited by M.\ 
Tarenghi and K.\ Kjar,
(ESO/ESA Workshop), p 3
\reference{} Taylor, C.\ L., Brinks, E., Pogge, R.\ W., \& Skillman,
E.\ D.\ 1994, AJ, 107, 971
\reference{} Thornley, M.\ D., \& Wilson, C.\ D.\ 1995, ApJ, 447, 616
\reference{} Tolstoy, E.\ 1996, PhD thesis, Rijksuniversiteit Groningen
\reference{} Toomre, A.\ 1964, ApJ, 139, 1217
\reference{} Toomre, A.\ 1981, in The Structure and Evolution of
Normal Galaxies, ed. S.\ M.\ Fall and D.\ Lynden-Bell (Cambridge:
Cambridge Univ.), p 111
\reference{} Tosi, M., Greggio, L., Marconi, G., \& Focardi, P.\ 1991, AJ, 102, 951
\reference{} Tully, R., Bottinelli, L., Fisher, J., Gouguenheim, L., Sancisi, R.,
\& van Woerden, H.\ 1978, A\&A, 63, 37
\reference{} Vader, J.P. \& Vigroux, L. 1991, A\&A, 246, 32
\reference{} van den Bergh, S.\ 1988, PASP, 100, 344
\reference{} van der Hulst, J.\ M., Skillman, E.\ D., Smith, T.\ R.,
Bothun, G.\ D., McGaugh, S.\ S., \& de Blok, W.\ J.\ G.\ 1993, AJ, 106, 548
\reference{} van Zee, L., Haynes, M.\ P., Salzer, J.\ J., \& Broeils,
A.\ H.\ 1996, AJ, 112, 129
\reference{} van Zee, L., Haynes, M.\ P., Salzer, J.\ J., \& Broeils,
A.\ H.\ 1997, AJ, 113, 1618
\reference{} Wilson, C.D., Scoville, N. \& Rice, W. 1991, AJ, 101, 129
\reference{} Yoshii, Y., \& Sommer-Larsen, J. 1989, MNRAS, 236, 779
\reference{} Young, L.\ M., \& Lo, K.\ Y.\ 1996, ApJ, 462, 203
\reference{} Young, J.\ S., \& Scoville, N.\ 1982, ApJ, 258, 467
\reference{} Zasov, A.V., \& Simakov, S.G. 1988, Astrophys, 29, 190
 
\end{references}
\end{document}